\documentclass[a4paper,11pt,usenatbib]{article}

\usepackage{jcappub}
\usepackage[T1]{fontenc}
\usepackage[utf8]{inputenc}
\usepackage{hyperref}
\usepackage{cprotect}
\usepackage{float}
\usepackage{subcaption}
\usepackage{caption}
\usepackage{times}
\usepackage{layouts}
\usepackage{booktabs}

\title{\boldmath Proca-stinated Cosmology I: A $N$-body code for the vector Galileon}

\author[a,b]{Christoph Becker}
\author[a]{, Christian Arnold}
\author[a]{, Baojiu Li}
\author[c]{ and Lavinia Heisenberg}

\affiliation[a]{Institute for Computational Cosmology, Department of Physics, Durham University, South Road, Durham DH1 3LE, United Kingdom}
\affiliation[b]{Institute for Data Science, Durham University, South Road, Durham DH1 3LE, United Kingdom}
\affiliation[c]{Institute for Theoretical Physics, ETH Zürich, Wolfgang-Pauli-Strasse 27, 8093, Zürich, Switzerland}

\emailAdd{christoph.becker@durham.ac.uk}

\abstract{We investigate the nonlinear growth of large-scale structure in the generalised Proca theory, in which a self-interacting massive vector field plays the role of driving the acceleration of the cosmic expansion. Focusing to the Proca Lagrangian at cubic order -- the cubic vector Galileon model -- we derive the simplified equations for gravity as well as the longitudinal and transverse modes of the vector field under the weak-field and quasi-static approximations, and implement them in a modified version of the {\tt ECOSMOG} $N$-body code. Our simulations incorporate the Vainshtein screening effect, which reconciles the fifth force propagated by the longitudinal mode of the cubic vector Galileon model with local tests of gravity. The results confirm that for all scales probed by the simulation, the transverse mode has a negligible impact on structure formation in a realistic cosmological setup. It is well known that in this model the strength of the fifth force is controlled by a free model parameter, which we denote as $\tilde{\beta}_3$. By running a suite of cosmological simulations for different values of $\tilde{\beta}_3$, we show that this parameter also determines the effectiveness of the Vainshtein screening. The model behaves identically to the cubic scalar Galileon for $\tilde{\beta}_3 \to 0$, in which the fifth force is strong in unscreened regions but is efficiently screened in high-density regions. In the opposite limit, $\tilde{\beta}_3 \to \infty$, the model approaches its `quintessence' counterpart, which has a vanishing fifth force but a modified expansion history compared to $\Lambda$CDM. This endows the model with rich phenomenology, which will be investigated in future works.}

\begin{document}
\def\lcdm{\Lambda{\rm CDM}}
\def\ecosmog{{\tt ECOSMOG} }

\maketitle
\flushbottom

\section{Introduction}
\label{sec:intro}

Our present understanding about the Universe is founded upon General Relativity (GR), which is the only theory that is compatible with the basic requisite of a single massless spin-2 field that respects Lorentz invariance \cite{Gupta:1954zz,Weinberg:1965rz,deRham:2014zqa}. Even though the predictions of GR have been validated against many tests, these tests are usually limited to small scales such as the solar system, and it leaves the cosmological scales underexplored \citep{Koyama:2015vza}. These latter scales coincide with those on which the dynamics of luminous matter within galaxies and at Mpc scales, as well as the expansion rate of the Universe, currently lack clear and convincing explanations. These enigma are commonly attributed to invisible energy contents that interact with gravity but not with baryons, called dark matter (motivated by e.g. galaxy dynamics) and dark energy (motivated by observed late time acceleration) \cite{Saunders:2017akc}. However, it is also possible that they are simply signatures that the law of gravity is modified on large scales, as exemplified by many modified gravity (MG) models \cite{Amendola:2015ksp,Heisenberg:2018vsk,Ferreira:2019xrr}.

The last decades have seen many attempts to modify GR. According to the Lovelock theorem, GR is the only theory with second-order local equations of motion for the metric field, which is derivable from a 4-dimensional action \cite{Koyama:2015vza}, and therefore modifications to GR often involve new dynamical degrees of freedom in addition to the metric field, non-locality, higher-dimensional spacetimes and/or higher-order equations. The simplest MG models, for example, usually involve a single scalar degree of freedom with self-interactions or interactions with curvature. It has been well-established that such models can be brought under the umbrella of the Horndeski theory \cite{Horndeski,Kobayashi:2011nu,Deffayet:2011gz}.

One of the well-known subclasses of the Horndeski theory is the 
Galileon model \citep{Nicolis:2008in, Deffayet:2009wt, Deffayet:2009mn}, 
a 4-dimensional effective theory which involves a scalar field with universal coupling to matter and derivative self-interactions. The theory implements the Vainshtein screening effect \cite{Vainshtein} -- a mechanism encountered in theories such as Fierz-Pauli massive gravity \cite{Babichev:2010jd} and the Dvali-Gabadadze-Porrati (DGP) model \cite{Dvali:2000hr} -- to decouple the scalar field from matter near massive objects and therefore can be compatible with Solar system tests of gravity. The model modifies the background expansion history such that it reaches a de Sitter solution in the future without requiring a cosmological constant. Its simplicity makes it possible to study its phenomenology with the help of cosmological $N$-body simulations \cite{Barreira:2013eea,Li:2013tda}.

In contrast to the scalar Galileon, the generalised Proca theory (GP) \cite{Heisenberg:2014rta,Allys:2015sht,Jimenez:2016isa}, involves a massive vector field, $A_{\mu}$, with a broken $U(1)$ gauge symmetry and second-order equation of motion (EOM). The theory features Galileon-type derivative self-interactions and couplings to matter. At the background level, the temporal component of the vector field, $A_0$, gives rise to a self-accelerating de Sitter attractor, corresponding to a dark energy equation of state $w_{\rm DE} = -1$ \cite{DeFelice:2016yws}. From the gravitational wave event GW170817 \cite{TheLIGOScientific:2017qsa} with accompanying gamma-ray burst GRB170817A \cite{Goldstein:2017mmi} and other optical counterparts, the speed of propagation of the gravitational waves $c_{T}$ has been tightly constrained to be identical to the speed of light, $c$. This places strong constraints on the allowed operators within the higher order GP Lagrangian. However, even with this restriction, the GP theory is still cosmologically interesting, with a theoretically consistent parameter space that is free of ghost and Laplacian instabilities \cite{DeFelice:2016yws}.

By introducing non-linear functions into the field Lagrangian of the GP theory to describe its derivative self interactions and couplings with matter, it is very versatile and flexible. However, in cosmological applications one often specialises to simple choices of these non-linear functions, such as power laws, and a number of studies have been conducted, leading to a good understanding of the cosmological behaviours of the model at background and linear levels. For example, in Ref.~\cite{deFelice:2017paw}, an MCMC likelihood analysis was performed for the particular GP theories proposed in Refs.~\cite{DeFelice:2016yws,DeFelice:2016uil}, by exploiting the observational data from type Ia supernovae (SNIa), cosmic microwave background (CMB), baryonic acoustic oscillations (BAO), the Hubble expansion rate $H(z)$, and redshift-space distortions (RSD). The cross correlation between galaxy field and the integrated Sachs Wolfe (ISW) effect, which has been a powerful probe to constrain the scalar Galileon models, has also been used to constrain parameters of the GP theory \cite{Nakamura:2018oyy}.


The aim of this paper is to carry on the analyses into the non-linear regime, beyond the use of linear perturbation theory \cite{DeFelice:2020sdq} or statistical field theory \cite{Heisenberg:2019ekf}, by using cosmological $N$-body simulations. From a phenomenological point of view, there are several reasons for doing so. One is that we know perturbation theory to not be good at quantifying the effects of screening, which is an inherently non-linear phenomenon. $N$-body simulations are the only known tool to accurately quantify the evolution of the Universe on small, highly non-linear, scales, and can be used to validate or calibrate the predictions of other approaches. Being able to probe small scales will enable us to test a given model against more observational data more accurately, e.g., access scales or regimes that are inaccessible to perturbation theory. To this end, we have developed a modified version of the \ecosmog code \cite{Li:2011vk}, which can be easily adapted to any variant of the GP theory. This is the first of a series of papers to explore the non-linear regime for this theory; here we will focus on deriving the simplified equations, code tests and initial simulations to gain some qualitative insight into its cosmological behaviour.

This paper is arranged as follows. In Section \ref{sec:proca_theory} we give a brief review of the key points of the GP theory, specialise to a simple variant of it, and derive the simplified Einstein and GP field equations of motion that are applicable to typical cosmological simulations which are featured by weak fields and slow motions of matter. 
A particularly detailed account will be given of the approximations used and their justifications. In Section \ref{sec:n_body_eqns} we introduce an internal unit system which is used to write the background and perturbation evolution equations into dimensionless forms. We give expressions of various physical quantities that are key to understanding the behaviours of the theory, and compare them with the predictions from other related gravity models. In Section \ref{sec:simulations_section}, we first carry out a range of tests of a new $N$-body code developed for simulating the GP field, and then show the first results from a suite of cosmological simulations. We show that the transverse mode of the GP vector field plays a negligible role in the non-linear evolution of the Universe, as it does in linear theory. We also demonstrate how the enhanced growth of non-linear cosmic structures and the screening of fifth force depends on the single additional parameter of the model. Finally we summarise, conclude and layout a future workplan in Section \ref{sect:discussion}.

Throughout this paper, we use the $(-,+,+,+)$ notation for the signature of the metric. We set $c=1$ except in expressions where $c$ appears explicitly. Greek indices run over $0,1,2,3$ while Roman indices run over $1,2,3$. $M_{\rm Pl}$ is the reduced Planck mass and is related to Newton's constant, $G$, by $M^{-2}_{\rm Pl}=8\pi G$.

\section{Generalised Proca Theory}
\label{sec:proca_theory}

This section gives a short description of the generalised Proca theory. We start from a complete form and then specialise to a particular case with a simple functional form of the Lagrangian. The choice of the theory and the resulting field equations are given in Section \ref{subsect:action}. In Section \ref{subsect:cosmo_eqns} we apply these to a perturbed spacetime around a flat homogeneous and isotropic Friedmann-Robertson-Walker (FRW) metric, to derive the equations which govern the dynamics of the Proca field and its effect on the total gravitational force. These will provide us with the essential equations for the $N$-body simulations.

\subsection{Action and general field equations}
\label{subsect:action}
In its standard form, the Proca action describes the dynamics of a massive vector field $A_{\mu}$, and is of little use to modify GR. This is because, since we observe no deviation from GR in our solar system, any new terms which we add to the Einstein-Hilbert action have to converge to GR in deep potentials. This requires the mass of the vector field to be so small, that it makes the field negligible on all scales. One way around this dilemma is by adding further terms to the action that make the behaviour of the vector field dependent on potential depths. This can be achieved by derivative self-interactions of the vector field. Interestingly, there exist only six derivative self-interactions that preserve the number of degrees of freedom of the vector field and do not create ghosts (such as the Ostrogradsky instability) \cite{Heisenberg:2014rta,Jimenez:2016isa}. The resulting four-dimensional action has the following structure \cite{Heisenberg:2014rta},
\begin{equation}\label{eq:proca_general}
    S = \int d^4x \sqrt{-g}\left[\mathcal{L}_m + \mathcal{L}_F + \displaystyle\sum_{i=2}^{5} \mathcal{L}_i\right],
\end{equation}
where $g$ denotes the determinant of the metric tensor $g_{\mu\nu}$, and $\mathcal{L}_m$ is the matter Lagrangian, which is related to the energy-momentum tensor of a perfect fluid as,
\begin{equation}\label{eq:lagrangian_matter}
    T^{(m)}_{\mu\nu} = -\frac{2}{\sqrt{-g}}\frac{\delta(\sqrt{-g}\mathcal{L}_m)}{\delta g^{\mu\nu}}.
\end{equation}
Assuming that matter is minimally coupled to gravity, $T^{(m)}_{\mu\nu}$ satisfies the standard conservation equation
\begin{equation}\label{eq:continuity_eq}
    \nabla^{\mu}T^{(m)}_{\mu\nu} = 0,
\end{equation}
where $\nabla^\mu$ denotes the covariant derivative compatible with $g_{\mu\nu}$. Introducing the first derivative of the vector field as $B_{\mu\nu}=\nabla_{\mu} A_{\nu}$, we can build the anti-symmetric Faraday tensor as $F_{\mu\nu} \equiv B_{\mu\nu} - B_{\nu\mu}$. The dynamics of $A_{\mu}$ is described by the kinetic term of the Proca Lagrangian, $\mathcal{L}_F$,
\begin{equation}\label{eq:proca_kinetic_lagrangian}
    \mathcal{L}_F = -\frac{1}{4}b_FF_{\mu\nu}F^{\mu\nu},
\end{equation}
and the self-interaction terms of the vector field, 
\begin{eqnarray}
    \mathcal{L}_2 &=& G_2(X, F_{\mu\nu}, \tilde{F}_{\mu\nu}),\\
    \mathcal{L}_3 &=& G_3(X)[B],\\
    \mathcal{L}_4 &=& G_4(X)R + G_{4,X}(X)([B]^2 - [B^2]),\\
    \mathcal{L}_5 &=& G_5(X)\mathcal{G}_{\mu\nu}B^{\mu\nu} - \frac{1}{6}G_{5,X}(X)([B]^3 - 3[B][B^2] + 2[B^3]) + \tilde{G}_5(X)\tilde{F}^{\alpha\mu}\tilde{F}^{\beta}_{\mu}B_{\alpha\beta},\\
    \mathcal{L}_6 &=& G_6(X)L^{\mu\nu\alpha\beta}B_{\mu\nu}B_{\alpha\beta} + \frac{1}{2}G_{6,X}(X)\tilde{F}^{\alpha\beta}\tilde{F}^{\mu\nu}B_{\alpha\mu}B_{\beta\nu},
\end{eqnarray}
where $X\equiv\frac{1}{2}A_\mu A^\mu$, $G_{2,3,4,5,6}$ are general algebraic functions of $X$, $\tilde{F} \equiv {}^{\ast}F$ is the Hodge-dual of the Maxwell tensor given by $\tilde{F}^{\mu\nu} = \mathcal{E}^{\mu\nu\alpha\beta}F_{\alpha\beta}/2$, where $\mathcal{E}^{\mu\nu\alpha\beta}$ is the Levi-Civita tensor satisfying the normalization $\mathcal{E}^{\mu\nu\alpha\beta}\mathcal{E}_{\mu\nu\alpha\beta}=-4!$. The square brackets around an operator designate the trace of a tensor. While $\mathcal{L}_{3,4,5,6}$ contain the derivative self interactions, the non-minimal derivative couplings of the vector field to the Ricci scalar $R$, the Einstein tensor $\mathcal{G}_{\mu\nu}$, and the double dual Riemann tensor $L^{\mu\nu\alpha\beta}$ defined by
\begin{equation}
    L^{\mu\nu\alpha\beta} = \frac{1}{4}\mathcal{E}^{\mu\nu\rho\sigma}\mathcal{E}^{\mu\nu\gamma\delta}R_{\rho\sigma\gamma\delta},
\end{equation}
where $R_{\rho\sigma\gamma\delta}$ is the Riemann tensor, are due to $\mathcal{L}_{4,5,6}$. Note that $b_F$ in Eq.~\eqref{eq:proca_kinetic_lagrangian} is a constant coefficient which has mass dimension zero in natural unit, and thus is sometimes set to $1$ in the literature; in physical unit it is not dimensionless, which is important when converting the field equations into code units, as will be seen below.

Exposing the full action given by Eq.~\eqref{eq:proca_general} to constraints from the observed gravitational wave event GW170817 \cite{TheLIGOScientific:2017qsa} with gamma-ray burst GRB170817A \cite{Goldstein:2017mmi} and other optical counterparts, we can already make a judgement on the viability of $\mathcal{L}_{4,5}$. The GW170817/GRB170817A event measured a speed of tensor perturbations $c_T$ very close to that of light $c$ \cite{Monitor:2017mdv}. In this work we consider the subclass of Proca theory with $\mathcal{L}_{5}=\mathcal{L}_6=0$ and $\mathcal{L}_4=\frac{1}{2}M^2_{\rm Pl}R$, which satisfies the requirement that  $c_T=c$\footnote{Note that this requirement does not exclude $\mathcal{L}_5$ entirely (as $\tilde{G}_5$ remains) and leaves interactions within $\mathcal{L}_6$ viable, since they are not sensitive to the background due to involved symmetries of the background and the interactions themselves \cite{Heisenberg:2018vsk}.}:
\begin{equation}\label{eq:lagrangian_cubic}
    \displaystyle\sum_{i=2}^{4} \mathcal{L}_i = G_2(X) + G_3(X)\nabla_{\mu}A^{\mu} + \left(\frac{M^2_{\rm Pl}}{2}\right)R,
\end{equation}
where $\mathcal{L}_{4}$ has simplified to the standard Einstein-Hilbert term. In the literature, a common choice of the functions $G_{2,3}(X)$ is the power-law form,
\begin{align}\label{eq:general_proca_algebra}
    G_{2}(X) = b_{2} X^{p_{2}}, & \quad & G_{3}(X) = b_{3} X^{p_{3}},
\end{align}
where $b_2 \equiv m^2$ is the mass-squared of the vector field that characterises the onset of the acceleration epoch, and $b_{3},p_2,p_3$ of mass dimension zero in natural unit. The choice is generic enough, leaving a viable parameter space in which the theory is free of ghost and Laplacian instabilities. Importantly, due to the derivative self-interaction of the vector field in $\mathcal{L}_3$, the gravitational effect of the field can be screened in dense regions as required by solar system tests. The screening mechanism in this model is known to be analogous to the Vainshtein mechanism for scalar Galileons \cite{DeFelice:2016cri}, as we will also demonstrate below, but there are also important differences between these two classes of models.

Based on the analyses of linear perturbations in this model, observational constraints on $p_{2,3}$ have been obtained in the literature, e.g., \cite{deFelice:2017paw,Nakamura:2018oyy,DeFelice:2020sdq}. In this work we set $p_{2} = p_3 = 1$ as a working example to study the qualitative behaviour of the Proca field and its impact on the cosmic structure formation, and leave the study of general functions $G_{2,3}(X)$ to future work. With this choice, the GP theory behaves as the standard scalar Galileon model in certain limits, as we will show later.

Having carefully chosen the components in our action, we can derive the EOM from it \cite{DeFelice:2016cri}. Variation with respect to $g^{\mu\nu}$ gives us the modified Einstein equation,
\begin{equation} \label{eq:gravity_eom}
\mathcal{G}_{\mu\nu}^{(F)} + \mathcal{G}_{\mu\nu}^{(2)} + \mathcal{G}_{\mu\nu}^{(3)} + \mathcal{G}^{(4)}_{\mu\nu} = \frac{1}{2}T_{\mu\nu}^{(m)},
\end{equation}
with
\begin{align}
\mathcal{G}_{\mu\nu}^{(F)} &= \frac{1}{4}g_{\mu\nu}(\nabla_{\rho}A_{\sigma}\nabla^{\rho}A^{\sigma} - \nabla_{\rho}A_{\sigma}\nabla^{\sigma}A^{\rho}) \\ \nonumber
    & \quad - \frac{1}{2} \left( \nabla_{\rho}A_{\mu}\nabla^{\rho}A_{\nu} + \nabla_{\mu}A_{\rho}\nabla_{\nu}A^{\rho} - 2\nabla_{\rho}A_{(\nu}\nabla_{\mu)}A^{\rho} \right) \\
\mathcal{G}_{\mu\nu}^{(2)} &= -\frac{1}{2}g_{\mu\nu}G_{2} - \frac{1}{2}G_{2,X}A_{\mu}A_{\nu} \\
\mathcal{G}_{\mu\nu}^{(3)} &= -\frac{1}{2}G_{3,X}(A_{\mu}A_{\nu}\nabla_{\rho}A^{\rho} + g_{\mu\nu}A^{\lambda}A_{\rho}\nabla_{\lambda}A^{\rho} - 2A_{\rho}A_{(\mu}\nabla_{\nu)}A^{\rho}) \\ 
\mathcal{G}^{(4)}_{\mu\nu} &= \frac{M_{Pl}^2}{2}\left(R_{\mu\nu} - \frac{1}{2}g_{\mu\nu}R\right).
\end{align}
where we have used the shorthand notation $G_{i,X} \equiv \partial G_i/\partial X$ with $i=2,3$. Variation with respect to $A_{\mu}$ gives us the EOM of the vector field,
\begin{equation} \label{eq:vector_eom}
    0 = \nabla_{\mu}F^{\mu\nu} - b_2A^{\nu} + 2b_3A^{[\mu}\nabla^{\nu]}A_{\mu},
\end{equation}
where the square bracket around indices mean their anti-symmetrisation. We can see from Eq.~\eqref{eq:gravity_eom}, that the existence of a vector field with derivative self-coupling induces additional gravitational interactions with matter. We want to study whether such interactions are viable in non-linear regimes of cosmological structure formation.

\subsection{Cosmological field equations}
\label{subsect:cosmo_eqns}
In order to derive the perturbation equations relevant for the study of large-scale structure formation, we work with the perturbed FRW metric in the Newtonian gauge
\begin{equation} \label{eq:FLRW_metric}
    g_{\mu\nu} = -(1+2\Psi){\rm d}t^2 + a^2(t)(1-2\Phi)\delta_{ij}{\rm d}x^i{\rm d}x^j,
\end{equation}
where $a(t)$ is the time-dependent scale factor which is normalised to $a(t_0)=1$ at the present day, and $\delta_{ij}=\text{diag}(+1,+1,+1)$ represents the spatial sector of the background metric that is taken here to be flat, $k=0$. 

We write the Proca field $A_\mu$ in its component form as $A_\mu=(A_0,A_i)=(\varphi,A_i)$, and further disentangle the spatial part of the Proca field, $A_i$, through the Helmholtz's theorem into a longitudinal and a transverse component
\begin{equation}\label{eq:helmholtz}
    A_i = B_i + \nabla_i \chi,
\end{equation}
where $B_i$ obeys the divergence-free condition, $\nabla^i B_i = 0$, and $\chi$ is the longitudinal scalar. Thus when deriving the components of the Einstein equations, we can apply the curl operator to filter out $B_i$ and the divergence operator to obtain the contribution of $\chi$.

Note that, rigorously speaking, the metric in Eq.~\eqref{eq:FLRW_metric} does not have enough physical degrees of freedom to fully describe the spacetime perturbations induced by a GP field. For example, the helicity-1 modes of the vector field produces vector mode perturbations of the metric. However, the interactions of the helicity-0 modes are typically stronger (in magnitude) than those of the helicity-1 modes \cite{deRham:2018qqo}. We will verify this numerically below, so that we can neglect their effects on cosmological structure formation\footnote{In the linear perturbation regime or for spherical mass distributions, it has been shown that the transverse component of the vector field vanishes identically, e.g. \cite{DeFelice:2016cri}.}. For this reason, our approach to treat the transverse component in this study is a `passive' one, where we solve $B_i$ as sourced by matter and $\chi$, but neglect the `backreaction' of $B_i$ on the evolution of the latter, with {\it a posteriori} check that such a neglecting is justified. This greatly simplifies the field equations solved in the $N$-body simulation, which would have been extremely cumbersome otherwise.

Solving cosmological structure formation is inherently computationally expensive, even without adding the transverse degree of freedom $B_i$ to the action. Therefore we apply two other approximations to further simplify the field equations. The first is the quasi-static approximation (QSA), under which all time derivatives of the field {\it perturbations} are assumed to be small compared with their spatial derivatives (e.g., $|\dot{\delta\varphi}|\ll|\delta\varphi_{,i}|$) and can therefore be dropped. We shall in addition assume that the time derivatives of the gravitational potentials are much smaller than their corresponding spatial derivatives,
\begin{equation}
    |\dot{\Phi}|\sim|\dot{\Psi}|\ll|\Phi_{,i}|\sim|\Psi_{,i}|,~~~\ddot{\Phi}\sim H\dot{\Phi}\ll|\Phi^{,i}_{\ ,i}|,
\end{equation}
where $_{,i}$ denotes derivative with respect to the comoving coordinate $x^i$ and an overdot the derivative with respect to the physical time $t$. As galaxy-survey data are still mostly available on scales small compared to the cosmological horizon, the QSA is usually a good approximation for $N$-body simulations. Nevertheless, we add the caveat here that for models like scalar Galileons and GP theory, the field equations are so complicated that a full $N$-body simulation in which all time derivatives are included is yet to be done, which means that the validity of the QSA remains largely an assumption. Actually, there have been suspicions that the approximations used to simplify the field equations in the scalar Galileon models, including QSA, may be linked to some artificial numerical issues encountered in simulations (see, e.g., \cite{Barreira:2013xea,Barreira:2013eea,Li:2013tda,Winther:2015pta} for some discussions). Due to this caveat, we shall explicitly mention it every time we apply the QSA.
The second is the weak-field limit (WFL), which says that terms such as $\varphi^{,i}\varphi_{,i}$ are much smaller compared with $\varphi^{,i}_{\ ,i}$. The application of both the QSA and the WFL considerably reduce the computational cost of running a simulation.

\subsubsection{The physical units of quantities}

Before proceeding to the cosmological field equations, and convert them into code-unit equations to be implemented in the $N$-body simulation code, it is useful to first clarify the physical unit of physical quantities in the GP theory.

Based on the action of the GP theory, we know that $G_2(X)$, $G_3(X)\nabla_\mu A^\mu$ and $G_4(X)R=c^4R/(16\pi G)$ have the same unit. Given that $[R]=L^{-2}$, $[c]=LT^{-1}$ and $[G]=M^{-1}L^3T^{-2}$, where $L,T,M$ represent respectively the units for length, time and mass, the unit of $G_4(X)R$ and hence of $G_2(X)$ and $G_3(X)\nabla_\mu A^\mu$, must be $ML^{-1}T^{-2}$. Therefore,
\begin{equation}
    ML^{-1}T^{-2} = [G_2(X)] = [b_2][X] = [b_2][A^\mu]^2,
\end{equation}
and 
\begin{equation}
    ML^{-1}T^{-2} = [G_3(X)\nabla_\mu A^\mu] = [b_3][L]^{-1}[A^\mu]^3
\end{equation}
where we have used $G_2(X)=b_2X$ and $G_3(X)=b_3X$. We choose the unit of the time component of the Proca field, $\varphi$, as $[\varphi]=L^{-1}$ so that the field has mass dimension 1 in natural unit as required (it is also possible to choose $[\varphi]=T^{-1}$ by rescaling $\varphi$ with $c$). Thus $[b_2] = MLT^{-2}$, $[b_3] = ML^3T^{-2}$ and similarly $[b_F] = ML^3T^{-2}$. Note that because $\varphi$ has the same unit as $A_i=B_i+\nabla_i\chi$, it follows that $\chi$ is dimensionless and $[B_i]=L^{-1}$.

\subsubsection{The modified Poisson equation}

The $(00)$ component of the perturbed Einstein equation, Eq.~\eqref{eq:gravity_eom}, after dropping terms according to the QSA and WFL, can be simplified as (with all $c$ factors restored)
\begin{eqnarray}\label{eq:perturbed_einstein}
    \frac{1}{2}\bar{\rho}_mc^2\left(1+\delta_m\right) &=& \frac{c^4}{16\pi G}\left[\frac{2}{a^2}\partial^2\Phi + 3\frac{H^2}{c^2}\right] - \frac{1}{4}b_Fa^{-4}\partial_iB_j\left(\partial^iB^j-\partial^jB^i\right)\nonumber\\
    && - \frac{1}{4}b_2\varphi^2 +\frac{1}{2}b_3\varphi^2\left[3\frac{H}{c}\varphi-a^{-2}\partial^2\chi\right].
\end{eqnarray}
Note that we replaced $\nabla$ by $\partial$ (which is the partial derivative with respect to the comoving coordinate) since $k = 0$, $\varphi = \bar{\varphi}(t) + \delta\varphi(t, \vec{x})$, where an overbar denotes background averaged quantities and $\delta\varphi$ the field perturbation; $\bar{\rho}_m$ and $\delta_m$ denote respectively the background density and density contrast of non-relativistic matter, where radiation has been neglected. We have, for this equation only, included the contribution from the transverse component of the Proca field (i.e., the term containing $B_i$), for illustration purpose, since it gives us a rough idea of what quantities to look at when comparing the contributions by the transverse versus longitudinal components to justify the neglecting of the former.

The above equation can be cleanly split into a purely background part, i.e., the modified Friedmann equation,
\begin{equation}\label{eq:modified_friedmann}
    3H^2 = 8\pi G\bar{\rho}_m(a) + \frac{1}{2}\beta_2{c^2}\bar{\varphi}^2 - 3\beta_3{c}H\bar{\varphi}^3,
\end{equation}
and a perturbation part, which corresponds to the modified Poisson equation (including the contribution from $B_i$ again)
\begin{equation}\label{eq:modified_poisson}
    \partial^2\Phi \approx \frac{4\pi G}{c^2}\bar{\rho}_ma^2\delta + \frac{1}{2}\beta_3\bar{\varphi}^2\partial^2\chi + \frac{1}{2}\beta_Fc^2a^{-4}\partial_iB_j\left(\partial^iB^j-\partial^jB^i\right),
\end{equation}
where we have redefined the parameters $b_2,b_3,b_F$ as $\beta_i \equiv 8\pi Gc^{-4}b_i$ with $i=2,3,F$. Note that $\beta_2$ is dimensionless while $[\beta_3]=[\beta_F]=L^{2}$.

Eq.~\eqref{eq:modified_poisson} solves the metric potential $\Phi$ provided a matter density field and configuration of $\chi$. However, it is the other potential $\Psi$ whose gradient is the gravitational force. The EOM of the $(ij)$ components of the perturbed Einstein equation contain further information on the relation between $\partial^2\Psi$, $A_i$, and matter perturbation, as well as between the sum of $\partial^2(\Phi+\Psi)$ and the anisotropic stress of the Proca field $A_i$. The latter can be used to solve $\Psi$ given $\Phi$. However, to the same approximation that the contribution from the transverse component $B_i$ is negligible to leading order, it can be shown that the anisotropic stress of the Proca field vanishes,
allowing us to approximate
\begin{equation}\label{eq:Phi_Psi}
    \Phi \approx - \Psi.
\end{equation}
In this case, $\chi$ behaves very similarly to the (cubic) scalar Galileon field. As a sanity check, we have confirmed that the expressions for $\mathcal{G}_{\mu\nu}^{(F,2,3)}$ we have found satisfy the Bianchi identity.

\subsubsection{Equation of motion for the longitudinal mode}

Proceeding with the EOM of the Proca field given in Eq.~\eqref{eq:vector_eom}, we begin with the temporal component, $\varphi$, which is given by,
\begin{equation}\label{eq:cvg_eom_0}
    0 = b_Fa^{-2}\left(\partial^2\dot{\chi}-c\partial^2\varphi\right) + b_2c\varphi - 3b_3H\varphi^2 + a^{-2}b_3c\varphi\partial^2\chi.
\end{equation}
The background part of this equation reads
\begin{equation}\label{eq:cvg_eom_0_eq1}
    b_2c = 3b_3H\bar{\varphi},
\end{equation}
which can be used to solve the background value of $\varphi$ given $H$. This can be further rewritten, using $\beta_2$ and $\beta_3$, as
\begin{equation}\label{eq:cvg_eom_0_eq1b}
    \beta_2c = 3\beta_3H\bar{\varphi}.
\end{equation}
On the other hand, at the perturbation level we have
\begin{equation}\label{eq:cvg_eom_0_eq2}
    b_F\left(c\partial^2\varphi - \partial^2\dot{\chi}\right) \approx b_3c\bar{\varphi}\partial^2\chi,
\end{equation}
where we have employed the WFL to neglect terms such as $b_2\delta\varphi$ and $-6b_3H\bar{\varphi}\delta\varphi$, and we have also used $\partial^2\varphi$ instead of $\partial^2\delta\varphi$ to lighten the notation. This equality makes it possible to replace the time derivatives of $\partial_i\chi$ and $\partial_i\varphi$ in the equation of motion for $\chi$. To see this, let us consider the EOM of the spatial component, $A_i = (\partial_i \chi, B_i)$,
\begin{eqnarray}\label{eq:cvg_eom_i}
    0 &=& b_F \left( c\partial^j\dot{\varphi} - \partial^j\ddot{\chi} - \ddot{B}^j \right) + b_F H \left(c\partial^j{\varphi} - \partial^j\dot{\chi} - \dot{B}^j\right) + b_F a^{-2}c^2\partial^2B^j\nonumber\\
    && - b_2c^2\left(\partial^j\chi + B^j\right) + b_3\left(c\dot{\bar{\varphi}} + 3cH \bar{\varphi} - a^{-2}c^2\partial^2\chi\right)\left(\partial^j\chi+B^j\right) - b_3c^2\bar{\varphi}\left(\partial^j\varphi-\bar{\varphi}\partial^j\Psi\right)\nonumber\\
    && - b_3c^2a^{-2}\left(\partial^j\partial^k\chi+\partial^j B^k\right)\left(\partial_k\chi+B_k\right).
\end{eqnarray}
We make two simplifications to this equation. First, as we are interested in the EOM for the longitudinal component $\chi$ in this subsection, we remove all the transverse components and leave them for the next subsection. Note that this does not mean that all terms involving $B_i$ should be dropped: for example, the term $B_k\partial^j B^k = \partial^j(B^k B_k )/2$ is a total derivative and has a nonzero divergence; on the other hand, terms such as $\ddot{B}^j$, $\dot{B}^j$ and $\partial^2B^j$ will be considered in the next subsection. Second, terms such as $B^j\partial^2\chi$, $\partial^jB^k\partial_k\chi$ and $B_k\partial^j\partial^k\chi$ are dropped on the ground that the `back-reaction' of $B^i$ on the dynamics of $\chi$ is negligible (the argument for this requires a better knowledge of the equation that governs $B^i$, and will be deferred to the next subsection).

Taking the divergence of Eq.~\eqref{eq:cvg_eom_i} to single out the longitudinal contributions, and dropping the terms that contain $B^i$, we find
 \begin{eqnarray}\label{eq:chi_eom_i_tmp1}
     0
     &=& b_F \left( c \partial^2\dot{\varphi} - \partial^2\ddot{\chi} \right) + b_{F}H \left(c\partial^2\varphi-\partial^2\dot{\chi}\right) - b_2c^2\partial^2\chi - b_3 c^2 \bar{\varphi} \left( \partial^2\varphi - \bar{\varphi}\partial^2 \Psi \right) \nonumber\\
     && + b_3 c \left( \dot{\bar{\varphi}} + 3H\bar{\varphi} \right) \partial^2\chi - b_3c^2a^{-2} \left[ \left( \partial^2\chi \right)^2 - \partial_i\partial_j\chi\partial^i\partial^j\chi \right],
\end{eqnarray}
This equation has two undesirable properties: first, it contains not just the spatial derivatives of $\chi$ but also of $\varphi$; second, it contains also spatial derivatives of $\dot{\chi}$ and $\ddot{\chi}$. On the face it seems to suggest that some sort of quasi-static approximation should be employed to drop terms such as $\partial^j \ddot{\chi}$ and $\partial^j \dot{\chi}$. It however turns out that one can use Eq.~\eqref{eq:cvg_eom_0_eq2} and its time derivative
\begin{equation}
    b_F \left( c \partial^2 \dot{\varphi} - \partial^2 \ddot{\chi} \right) = b_3 c \dot{\bar{\varphi}} \partial^2 \chi + b_3 c \bar{\varphi} \partial^2 \dot{\chi},
\end{equation}
to rewrite Eq.~\eqref{eq:chi_eom_i_tmp1} in the following more convenient form,
\begin{equation}\label{eq:chi_eom_i1}
    \left[\frac{b_2}{b_3}-2c^{-1}\left(\dot{\bar{\varphi}}+2H\bar{\varphi}\right)+\frac{b_3}{b_F}\bar{\varphi}^2\right]\partial^2\chi + a^{-2}\left[\left(\partial^2\chi\right)^2-\partial_i\partial_j\chi\partial^i\partial^j\chi\right] = \bar{\varphi}^2\partial^2\Psi.
\end{equation}
Note that this means all time derivatives are eliminated exactly, so that we do not have to resort to the QSA. As a final step, we replace $b_{2,3,F}$ with $\beta_{2,3,F}$ as before, and use the modified Poisson equation, \eqref{eq:modified_poisson} (excluding the contributions from $B^i$) and the relation $\Phi\approx-\Psi$ in Eq.~\eqref{eq:Phi_Psi} to eliminate $\Psi$, and obtain
\begin{eqnarray}\label{eq:chi_eom_i}
    \left[\frac{1}{\bar{\varphi}^2}\frac{\beta_2}{\beta_3}-2c^{-1}\left(\frac{\dot{\bar{\varphi}}}{\bar{\varphi}^2}+2H\frac{1}{\bar{\varphi}}\right)+\frac{\beta_3}{\beta_F}-\frac{1}{2}\beta_{3}\bar{\varphi}^2\right]\partial^2\chi && \nonumber\\ 
    + \frac{1}{\bar{\varphi}^2a^{2}}\left[\left(\partial^2\chi\right)^2-\partial_i\partial_j\chi\partial^i\partial^j\chi\right] &=& \frac{4\pi G}{c^2} a^2\bar{\rho}_m\delta_m = \nabla^2\Phi_N,
\end{eqnarray}
where $\Phi_N=\Psi_N$ is the standard Newtonian potential. This is the main equation that we will convert to code unit and implement into the $N$-body simulation code in the next section.

\subsubsection{Equation of motion for the transverse mode}
\label{subsect:trans_eom}

Singling out the transverse part of Eq.~\eqref{eq:cvg_eom_i} by applying the curl operator once would leave a numerically inconvenient equation behind. This can be bypassed by simply applying the curl once more on itself and simplifying things using the vector identity,
\begin{equation}
    \nabla\times\left(\nabla\times{\bf B}\right) = \nabla\left(\nabla\cdot{\bf B}\right) - \nabla^2{\bf B} = -\nabla^2{\bf B},
\end{equation}
where in the second step we have used the fact that $\bf B$ satisfies $\nabla\cdot{\bf B}=0$. Thus we obtain, for the EOM of $B_i$,
\begin{eqnarray} \label{eq:B_eom_i_complex}
    0 &=& b_Fc^{-2}\partial^2\ddot{\bf B} + b_Fc^{-2}H\partial^2\dot{\bf B} -a^{-2}b_F\partial^4{\bf B} + b_2\partial^2{\bf B} - b_3\left(c^{-1}\dot{\bar{\varphi}}+3c^{-1}H\bar{\varphi}-a^{-2}\partial^2\chi\right)\partial^2{\bf B}\nonumber\\
    && + b_3\left[\partial^2\Psi\vec\partial\varphi^2 - \partial^2\varphi^2\vec\partial\Psi + \left({\vec\partial}\Psi\cdot{\vec\partial}\right)\vec\partial\varphi^2 - \left({\vec\partial}\varphi^2\cdot{\vec\partial}\right)\vec\partial\Psi\right]\nonumber\\
    && + b_3\partial^2\chi{\vec\partial}\left(c^{-1}\dot{\varphi}+3c^{-1}H\varphi-a^{-2}\partial^2\chi\right) - b_3{\partial}^2\left(c^{-1}\dot{\varphi}+3c^{-1}H\varphi-a^{-2}\partial^2\chi\right)\left({\vec\partial}\chi+{\bf B}\right)\nonumber\\
    && + b_3\left[\left({\vec\partial}\chi+{\bf B}\right)\cdot{\vec\partial}\right]{\vec\partial}\left(c^{-1}\dot{\varphi}+3c^{-1}H\varphi-a^{-2}\partial^2\chi\right)\nonumber\\
    && - b_3\left[{\vec\partial}\left(c^{-1}\dot{\varphi}+3c^{-1}H\varphi-a^{-2}\partial^2\chi\right)\cdot{\vec\partial}\right]\left({\vec\partial}\chi+{\bf B}\right),
\end{eqnarray}
where we have used $\vec\partial$ to denote the vector gradient.  This expression is still too complex for a cosmological simulation, making it necessary to apply further simplifications with the following arguments. 

First, the QSA is applied to drop the time derivatives of $B^i$, namely $|\partial^2\ddot{\bf B}|\sim|H\partial^2\dot{\bf B}|\ll|\partial^4{\bf B}|$, from the equation. Therefore, the above equation can be considered as a constraint equation in which ${\bf B}$, or $\partial^2{\bf B}$, is sourced by various terms. The term $a^{-2}b_F\partial^4{\bf B}$ contains the Laplacian of $\partial^2{\bf B}$, which should be what other terms are compared against to decide the relative importance.

For example, we start with comparing the magnitudes of $b_2\partial^2{\bf B}$ and $a^{-2}b_F\partial^4{\bf B}$. Schematically we can write $|\partial^2{\bf B}|\sim\eta^{-2}|\Delta{\bf B}|$, where $\eta$ is the size (in Mpc$/h$) of the mesh cells on which we will discretise the equation and numerically solve it in the simulation, and $\Delta{\bf B}$ is the typical difference between the values of ${\bf B}$ in neighbouring cells of the mesh. Likewise, $|\partial^4{\bf B}|\sim\eta^{-4}|\Delta{\bf B}|$\footnote{Because we are only interested in an order-of-magnitude estimate, we neglect the fact that the $\Delta{\bf B}$ values are different in these two cases, and assume that they are of similar magnitudes.}. Therefore, the ratio of these two quantities can be estimated as
\begin{equation}\label{eq:xx1}
    \frac{|b_2\partial^2{\bf B}|}{|a^{-2}b_F\partial^4{\bf B}|} \sim \frac{b_2}{b_F}\eta^2 = \frac{\beta_2}{\beta_F}\eta^2 = \frac{\tilde{\beta}_2}{\tilde{\beta}_F}\eta^2\left(\frac{c}{H_0}\right)^{-2}
\end{equation}
where we have defined the dimensionless variables 
\begin{eqnarray}\label{eq:beta_tildes}
    \tilde{\beta}_2 &\equiv& \beta_2,\nonumber\\
    \tilde{\beta}_{3,F} &\equiv& \beta_{3,F}\left(\frac{c}{H_0}\right)^{-2},
\end{eqnarray}
that will be used later to write the field equations in code unit. We have the freedom to set $\tilde{\beta}_F=1$ by a field redefinition, $\tilde{\beta}_3$ is a free parameter of the model studied here for which we are interested in $\mathcal{O}(10^{-6})\lesssim\tilde{\beta}_3\lesssim\mathcal{O}(100)$, and $\tilde{\beta}_2$ is related to $\tilde{\beta}_3$ through Eq.~\eqref{eq:omega_cvg} below as $\tilde{\beta}_2 = - 54^{1/3}(1-\Omega_m)^{1/3}\tilde{\beta}_3^{2/3}$ with $\Omega_m\approx0.3$ being the matter density parameter today. Therefore $\tilde{\beta}_2\lesssim70$; combining with the fact that $c/H_0\approx3000h^{-1}$Mpc and $\eta\lesssim1h^{-1}$Mpc in typical simulations, this means that the ratio in Eq.~\eqref{eq:xx1} is much smaller than 1, and so the term $b_2\partial^2{\bf B}$ can be neglected from Eq.~\eqref{eq:B_eom_i_complex}.

As another example, we compare $b_3c^{-1}\dot{\bar{\varphi}}\partial^2{\bf B}$ and $b_3c^{-1}H\bar{\varphi}\partial^2{\bf B}$ against $a^{-2}b_F\partial^4{\bf B}$. We can regard the former two quantities as the same order because $\dot{\bar{\varphi}}\sim H\bar{\varphi}$, so we focus on $b_3c^{-1}H\bar{\varphi}\partial^2{\bf B}$. The ratio is 
\begin{equation}\label{eq:xx2}
    \frac{|b_3c^{-1}H\bar{\varphi}\partial^2{\bf B}|}{|a^{-2}b_F\partial^4{\bf B}|} \sim \frac{b_3}{b_F}c^{-1}H\bar{\varphi}\eta^2 = \frac{\tilde{\beta}_3}{\tilde{\beta}_F}c^{-1}H\bar{\varphi}\eta^2 = \frac{\tilde{\beta}_2}{3\tilde{\beta}_F}\eta^2\left(\frac{c}{H_0}\right)^{-2} \ll 1,
\end{equation}
where in the last equality we have used Eq.~\eqref{eq:cvg_eom_0_eq1b}. Therefore these terms can also be dropped from Eq.~\eqref{eq:B_eom_i_complex}.

Second, consider the term $b_3a^{-2}\partial^2\chi\partial^2{\bf B}$. We have for cosmological objects $|\Phi_{\text{N}}| \lesssim \mathcal{O}(10^{-4})$, and can use Eq.~\eqref{eq:chi_eom_i} to estimate the size of $\chi$. This can be divided into two cases. The first is when the left-hand side of Eq.~\eqref{eq:chi_eom_i} is dominated by the first term, which is linear in $\partial^2\chi$ -- there are four terms in the bracket in front of $\partial^2\chi$ in Eq.~\eqref{eq:chi_eom_i}, and with a lengthy but trivial calculation it can be shown that their relative magnitudes vary depending on the parameter value of $\tilde{\beta}_3$ and the time $a$. For simplicity, this whole bracket can be written as $\epsilon\left(\tilde{\beta}_3/\tilde{\beta}_2\right)+\left(\tilde{\beta}_3/\tilde{\beta}_F\right)$, where $\epsilon$ is a time-dependent function of order $\mathcal{O}(10)$ or larger. In the second case, the non-linear term dominates the left-hand side of Eq.~\eqref{eq:chi_eom_i}, and one has 
\begin{equation}\label{eq:yy1}
    \frac{1}{a^2\bar{\varphi}^2}\frac{1}{\eta^4}|\Delta\chi|^2 \sim \frac{1}{\eta^2}|\Delta\Phi_N| \Rightarrow \frac{\tilde{\beta}_3}{\tilde{\beta}_2}|\Delta\chi| \sim \frac{1}{3}a\sqrt{|\Phi_N|}\eta\left(\frac{c}{H_0}\right)^{-1}\frac{H_0}{H},
\end{equation}
where we have used
\begin{equation}\label{eq:yy2}
    \bar{\varphi}^{-1} = \frac{3\tilde{\beta}_3}{\tilde{\beta}_2}\frac{c}{H_0}\frac{H}{H_0},
\end{equation}
which itself is derived from Eq.~\eqref{eq:cvg_eom_0_eq1b}. Following the previous logic, the ratio to $b_Fa^{-2}\partial^4{\bf B}$ is given by
\begin{equation}\label{eq:xx3}
    \frac{|b_3a^{-2}\partial^2\chi\partial^2{\bf B}|}{|b_Fa^{-2}\partial^4{\bf B}|} \sim \frac{b_3}{b_F}\frac{\eta^{-2}|\Delta\chi|\cdot\eta^{-2}|\Delta{\bf B}|}{\eta^{-4}|\Delta{\bf B}|} \sim \frac{\tilde{\beta}_3}{\tilde{\beta}_F}|\Delta\chi|.
\end{equation}
It can then be straightforwardly checked that the ratio in Eq.~\eqref{eq:xx3} is always much smaller than 1 for both cases, and when either $\tilde{\beta}_3/\tilde{\beta}_2$ or $\tilde{\beta}_3/\tilde{\beta}_F$ dominates in the first case. Therefore this term can also be dropped from Eq.~\eqref{eq:B_eom_i_complex}. 

Third, consider the terms such as $\partial^2\Psi\vec\partial\varphi^2 \sim \bar{\varphi}\partial^2\Psi\vec\partial\varphi$ that can also source $\partial^2{\bf B}$, in the second line of Eq.~\eqref{eq:B_eom_i_complex}. Integrating Eq.~\eqref{eq:cvg_eom_0_eq2} once, one finds 
\begin{equation}\label{eq:tmp3}
    b_F\left(c\partial_i\varphi-\partial_i\dot{\chi}\right) \approx b_3c\bar{\varphi}\partial_i\chi \Rightarrow \partial_i\varphi \approx c^{-1}\partial_i\dot{\chi} + \frac{b_3}{b_F}\bar{\varphi}\partial_i\chi,
\end{equation}
so that $|\partial_i\varphi|$ is approximately of the same order as $b_3/b_F\bar{\varphi}|\partial_i\chi|$ or $c^{-1}|\partial\dot{\chi}|\sim c^{-1}H|\partial_i\chi|$, whichever dominates. In practice, the two terms on the right-hand side of Eq.~\eqref{eq:tmp3} can differ by a factor of up to $\mathcal{O}(10)$. To demonstrate that terms such as $b_3\bar{\varphi}\partial^2\Psi\vec\partial\varphi$, instead of showing that its amplitude is much smaller than $|b_Fa^{-2}\partial^4{\bf B}|$, we will seek to show that it is much smaller than the amplitude of certain other terms in Eq.~\eqref{eq:B_eom_i_complex}, in particular $b_3\partial^2\chi\vec\partial^i\partial^2\chi$ -- consider the ratio
\begin{eqnarray}\label{eq:xx4}
    \frac{|b_3\bar{\varphi}\partial^2\Psi\vec\partial\varphi|}{|b_3\partial^2\chi\vec\partial\partial^2\chi|} \sim c^{-1}H\bar{\varphi}\frac{|\partial^2\Phi_N|\cdot|\vec\partial\chi|}{|\partial^2\chi\vec\partial\partial^2\chi|} \sim c^{-1}H\bar{\varphi}\frac{\eta^{-3}|\Delta\chi|^2}{\eta^{-5}|\Delta\chi|^2}\left[\epsilon\frac{\tilde{\beta}_3}{\tilde{\beta}_2}+\frac{\tilde{\beta}_3}{\tilde{\beta}_F}\right] && \nonumber\\
    \sim \frac{\tilde{\beta}_2}{3\tilde{\beta}_3}\left[\epsilon\frac{\tilde{\beta}_3}{\tilde{\beta}_2}+\frac{\tilde{\beta}_3}{\tilde{\beta}_F}\right]\eta^2\left(\frac{c}{H_0}\right)^{-2} &\ll& 1,
\end{eqnarray}
where in the first `$\sim$' we have assumed that $|\vec\partial\varphi|\sim c^{-1}H|\vec\partial\chi|\gg (b_3/b_F)\bar{\varphi}|\vec\partial\chi|$, $|\Phi_N|\sim|\Phi|$, and in the second `$\sim$' we have assumed that the term proportional to $\nabla^2\chi$ dominates the left-hand side of Eq.~\eqref{eq:chi_eom_i}. It can be similarly shown that the ratio is also much smaller than 1 in the other limits, e.g., when $|\vec\partial\varphi|\sim (b_3/b_F)\bar{\varphi}|\vec\partial\chi| \gg c^{-1}H|\vec\partial\chi|$ and/or the non-linear term dominates the left-hand side of Eq.~\eqref{eq:chi_eom_i}, though the details are omitted here for brevity. This indicates that these source terms can also be safely dropped off from Eq.~\eqref{eq:B_eom_i_complex}.

Fourth, in Eq.~\eqref{eq:B_eom_i_complex} a number of terms can be neglected by realising that $c^{-1}|\vec\partial\dot\varphi|\sim c^{-1}H|\vec\partial\varphi|\ll a^{-2}|\vec\partial\partial^2\chi|$. The proof of these relations is straightforward and we shall not repeat them here.
 
Finally, therefore, we see that terms like $\partial^2\chi\partial_i\partial^2\chi$ are the remaining sources for $\partial^4B^i$. For the former, we have $|\partial^2\chi\partial_i\partial^2\chi|\sim\eta^{-5}|\chi|\cdot|\Delta\chi|$, and for the latter we have $|\partial^4B^i|\sim\eta^{-4}|\Delta B^i|$. This suggests that $|\Delta B^i|\sim|\chi|\cdot\eta^{-1}\Delta\chi\ll\eta^{-1}|\Delta\chi|$ and confirms that $|\partial B^i|\ll|\partial^2\chi|$ is self-consistent.

With the above approximations, the equation can be simplified to
\begin{equation} \label{eq:B_eom_i}
    \beta_F\partial^4B^i = \beta_3\partial^j\left[\partial_i\chi\partial_j\partial^2\chi - \partial_j\chi\partial_i\partial^2\chi\right].
\end{equation}

Eqs.~(\ref{eq:modified_poisson}, \ref{eq:chi_eom_i}, \ref{eq:B_eom_i}) are the key equations of this paper -- the last one is used to calculate $B^i$ and verify that the transverse component makes negligible contribution (`feedback') to the dynamics of $\Phi$ and $\chi$, the second one is used to solve $\chi$ given a matter distribution, and finally the first one is used to find the total gravitational potential (and therefore the total gravity force) for the given matter distribution and the resulting spatial configuration of $\chi$.

As $\tilde{\beta}_{3} \equiv b_3(8\pi GH_0^2)/(c^6)$ is the only `free' parameter that enters in all three key equations it is practical to use it as the model parameter. Previous works denote the model parameters that behave similarly to $\tilde{\beta}_{3}$ as $\lambda$ \cite{deFelice:2017paw,Nakamura:2018oyy} and $q_v$ \cite{DeFelice:2016uil,Heisenberg:2019ekf}, which are both inversely proportional to $\tilde{\beta}_{3}$. We do not present the exact relations between those parameters and $\tilde{\beta}_3$ here.

\section{$N$-body Equations}
\label{sec:n_body_eqns}

In this section we describe the numerical implementation of the above-derived equations into the $N$-body code \ecosmog \cite{Li:2011vk}. For this purpose, we will need to recast the equations in {\tt ECOSMOG}'s code units, in which all quantities are rescaled so that only dimensionless quantities appear. In order to acquire a better understand about the cvG model behaviour we juxtapose it with the well studied cosmologies of $\lcdm$, self-accelerating branch of the Dvali-Gabadadze-Porrati model (sDGP, \cite{Dvali:2000hr}), and the tracker solution of the cubic scalar Galileon (csG, \cite{Barreira:2013xea,Barreira:2014ija})\footnote{Note that although the csG is a generalisation of the sDGP that arises from its decoupling limit, their phenomenology is very different.} where appropriate. For the csG model, we assume that for the entire time period of interest here the model follows the {\it tracker solution} \cite{DeFelice:2010pv}, which is an attractor of the evolution; in practice, the time at which the model merges onto this common late-time evolution trajectory depends on the initial conditions of the background csG field, but it was demonstrated in Ref.~\cite{Barreira:2014jha} that the merging onto the tracker solution should happen before the onset of the acceleration era, $a \sim 0.5$, in order to satisfy CMB constraints. In all visualisations of the models we adopt the following two cosmological parameters: $\Omega_m=0.3089$ and $H_0=67.74$ kms/s/Mpc. For the sDGP specific parameters we use $\Omega_{rc}=0.25$, while the csG specific parameters are the following: $\Omega_{\varphi} = 1-\Omega_m$, $\xi = \sqrt{6\Omega_{\varphi}}$, $c_2 = -1$, and $c_3 = 1/(6\xi)$ (see Ref.~\cite{Barreira:2014ija} for more details). To better understand the effects of the fifth force we compare the csG and cvG model to their quintessence counterpart, QCDM, which is a variant that only considers the modified background expansion history, but uses standard Newtonian gravity, in the simulation.

\subsection{Code units}
\label{sec:code_units}

In order to implement the equations into {\tt ECOSMOG}, we introduce a set of dimensionless quantities that are based on $H_0^{-1}$ for measuring time, the simulation box size $L$ in units of Mpc$/h$, the particle velocity $v$, the critical density today $\rho_{c0} = 8\pi G/(3H_0^2)$ and the matter density $\Omega_{m} = \Omega_{b} + \Omega_{c}$ at the present day:
\begin{align}\label{eq:code_unit}
    \tilde{x} = \frac{x}{L} & \quad & \tilde{\rho} = \frac{\rho a^3}{\rho_{\text{c0}}\Omega_{\text{m}}} & \quad & \tilde{v} = \left(\frac{a}{LH_0}\right)v \nonumber\\
    \tilde{c} = \frac{c}{H_0L} & \quad & d\tilde{t} = H_0a^{-2}{dt} & \quad & \tilde{\Phi} = \left(\frac{ac}{LH_0}\right)^2\Phi.
\end{align}
Notice that, in order to simplify the equations in code units, we have introduced the super-comoving coordinate time $\tilde{t}$  \cite{Martel:1997hk}. All quantities that we in the super-comoving system are from now on marked by a tilde. In this coordinate system the background matter density is unity, $\tilde{\bar{\rho}} = 1$. 

To transform the quantities introduced by the Proca theory to code units, we need to know their physical units. As mentioned above, the Proca field $A_\mu$ has mass dimension 1 in natural unit, and we have $[A_i] = L^{-1}$ in physical unit, so that the longitudinal mode $\chi$ is dimensionless. However, since $\chi$ plays an equivalent role as $\Phi$ in determining the force, we transform it into code unit in the same way as for $\Phi$; on the other hand, the transverse component $B_i$ has unit $[B_i]=L^{-1}$, we multiply it by the box size $L$ to get $\tilde{B}_i$; to get the code-unit expression for $\bar{\varphi}$, which has unit $[\bar{\varphi}]=L^{-1}$, instead of multiplying it by $L$, we multiply it by $c/H_0$ because this variable is only used to calculate background quantities. The results are:
\begin{align}\label{eq:code_unit_extradof}
    \tilde{\varphi} = \frac{c}{H_0}\bar{\varphi} & \quad & 
    \tilde{B}_i = a^4\tilde{c}^4LB_i & \quad & \tilde{\chi} = \left(\frac{ac}{LH_0}\right)^2\chi,
\end{align}
where we have also included a factor $a^4\tilde{c}^4$ in $\tilde{B}_i$ to further simply the code-unit equation of Eq.~\eqref{eq:B_eom_i}.

\subsection{Background and perturbation equations}
The modified Friedmann equation, Eq.~\eqref{eq:modified_friedmann}, can be simplified as
\begin{equation}
    3H^2 = 8\pi G\bar{\rho}_m(a) - \frac{1}{18}\frac{\tilde{\beta}^3_2}{\tilde{\beta}^2_3}\frac{H_0^4}{H^2},
\end{equation}
where we have used Eq.~\eqref{eq:cvg_eom_0_eq1b} and the definitions of $\tilde{\beta}_{2}$ and $\tilde{\beta}_{3}$.

\begin{figure}[tbp]
    \centering 
    \includegraphics[width=.98\textwidth]{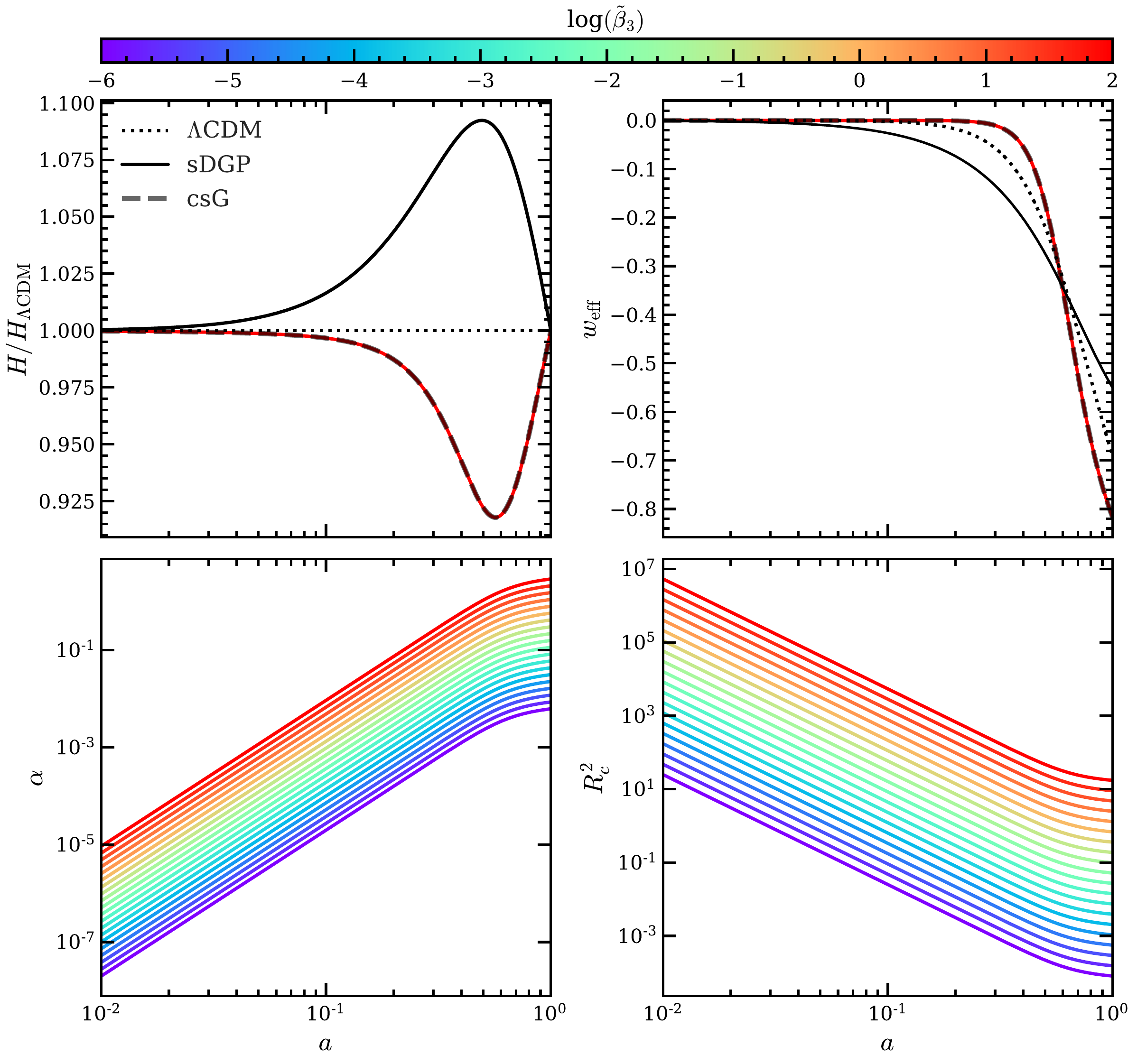}
     \caption{The time evolution of various background quantities in the cvG (coloured lines with different $\tilde{\beta}_3$ values as indicated by the colour bar on the top), csG (grey dashed line), sDGP (black solid line) and $\lcdm$ (black dotted line) models. {\it Top left:} The ratios of Hubble expansion rate in other cosmologies with respect to $\lcdm$; note that the cvG results do not depend on $\tilde{\beta}_3$ and are identical to the csG prediction. {\it Top right:} The effective 
     equation of state parameter, $w_{\rm eff}$, given in Eqs.~\eqref{eq:w_eff}. {\it Bottom left:} The cvG model function $\alpha$ given in Eq.~\eqref{eq:func_alpha}. {\it Bottom right:} The cvG model function $R_c^2$ given in Eq.~\eqref{eq:func_Rc2}.
     } 
     \label{fig:background_factors}
\end{figure}

As the Friedmann equation is commonly expressed as a relation between density parameters today, we can follow this practice for the Proca field by defining $\Omega_{P}$ (similar to $\Omega_\varphi$ in the csG model) as the links between the two coupling constants $\tilde{\beta}_2$ and $\tilde{\beta}_3$ through
\begin{equation}\label{eq:omega_cvg}
    \Omega_{P} \equiv -\frac{1}{54} \frac{\tilde{\beta}^3_2}{\tilde{\beta}^2_3} = 1-\Omega_{m},
\end{equation}
where note that $\tilde{\beta}_2<0$. This leads to the following result of $E(a)\equiv H(a)/H_0$ for the cvG model, which we show together with the expressions for the other cosmologies for clarity,
\begin{equation}\label{eq:friedmann_code}
    E^2 = 
    \left \{
    \def\arraystretch{2.2}
    \begin{array}{ll}
        \Omega_{m}a^{-3} + \Omega_{\Lambda},  & \Lambda{\rm CDM}, \, \\
        \Omega_{m}a^{-3} + 2\Omega_{{rc}} + 2\sqrt{\Omega_{m}a^{-3} + 2\Omega_{{rc}}},  & {\rm sDGP}, \, \\
        \frac{1}{2}\left[\Omega_{m}a^{-3} + \sqrt{\Omega^2_{m}a^{-6} + 4\left(1-\Omega_{m}\right)}\right],  & {\rm cvG},\, {\rm csG},\, {\rm QCDM}. \,
    \end{array}
    \right.
\end{equation}
where we have assumed the Universe to be spatially flat ($k=0$) and considered only non-relativistic matter; the inclusion of radiation and massive neutrinos is straightforward. Therefore, the background expansion history in this model is completely determined by $H_0$ and $\Omega_m$, and mimics precisely tracker solution of the csG model, e.g., \cite{Barreira:2013xea,Barreira:2014ija}. This can be seen clearly in the top-left panel of Fig.~\ref{fig:background_factors}, which shows a comparison of the background expansion history in the cvG model with those of the DGP and csG models.

We also give the effective equation of state, $w_{\rm eff} = -1 - 2\dot{H}/(3H^2)$, in the top-right panel of the same figure
\begin{equation}\label{eq:w_eff}
    w_{\rm eff} = 
    \left \{
    \def\arraystretch{2.2}
    \begin{array}{ll}
        \frac{\Omega_{m} - 1}{1 - \Omega_{m} + \Omega_{m}a^{-3}},  & \Lambda{\rm CDM}, \, \\
        \frac{2}{3\beta_{\rm sDGP}},  & {\rm sDGP}, \, \\
        -1 + \frac{\Omega_{m}a^{-3} + \frac{\Omega^2_{m}a^{-6}}{\sqrt{\Omega^2_{m}a^{-6} + 4(1-\Omega_m)}}}{\Omega_{m}a^{-3} + \sqrt{\Omega^2_{m}a^{-6} + 4(1-\Omega_{m})}},  & {\rm cvG},\, {\rm csG},\, {\rm QCDM}. \,
    \end{array}
    \right.
\end{equation}

The modified Poisson equation, Eq.~\eqref{eq:modified_poisson}, takes the following form in code unit,
\begin{equation}\label{eq:poisson_code}
    \tilde{\partial}^2\tilde{\Phi} = \frac{3}{2}\Omega_{m}a\left(\tilde{\rho}-1\right) + \alpha\tilde{\partial}^2\tilde{\chi},
\end{equation}
where 
\begin{equation}\label{eq:func_alpha}
    \alpha(a) \equiv \frac{1}{2}\tilde{\beta}_3\tilde{\varphi}^2 = \frac{1}{2\sqrt[3]{2}}\tilde{\beta}_3^{1/3}\Omega_P^{-1/3}\left[\sqrt{\Omega_m^2a^{-6}+4\Omega_P}-\Omega_ma^{-3}\right],
\end{equation}
is a time-dependent function that is fully fixed by specifying $\Omega_m$ and $\tilde{\beta}_3$. The left-bottom panel of Fig.~\ref{fig:background_factors} shows how $\alpha(a)$ evolves in time for different values of $\tilde{\beta}_3$.

Recasting Eq.~\eqref{eq:B_eom_i} in code units gives
\begin{equation}\label{eq:B_eom_i_code}
    \tilde{\partial}^4\tilde{B}^i = \frac{\tilde{\beta}_3}{\tilde{\beta}_F}\tilde{\partial}^j\left[\tilde{\partial}_i\tilde{\chi}\tilde{\partial}_j\tilde{\partial}^2\tilde{\chi} - \tilde{\partial}_j\tilde{\chi}\tilde{\partial}_i\tilde{\partial}^2\tilde{\chi}\right].
\end{equation}
As mentioned above, we can set $\tilde{\beta}_F=1$, which is achievable by a field redefinition, without loss of generality. Therefore $\tilde{\beta}_F$ is not a free parameter of the Proca model here.

Finally, the EOM for the longitudinal mode of the Proca field, $\chi$, Eq.~\eqref{eq:chi_eom_i}, can be rewritten in code unit as,
\begin{eqnarray}\label{eq:poisson_eom_code}
    \frac{3}{2}\Omega_{m}a\left(\tilde{\rho}-1\right)
    &=& \left[\frac{\tilde{\beta}_2}{\tilde{\beta}_3\tilde{\varphi}^2}-6\frac{\tilde{\beta}_3}{\tilde{\beta}_2}\left(2-\frac{H'}{H}\right)E^2+\frac{\tilde{\beta}_3}{\tilde{\beta}_F}-\frac{1}{2}\tilde{\beta}_{3}\tilde{\varphi}^2\right]\tilde{\partial}^2\tilde{\chi} \nonumber\\
    && + \frac{1}{\tilde{\varphi}^2a^{4}}\left[\left(\tilde{\partial}^2\tilde{\chi}\right)^2-\tilde{\partial}_i\tilde{\partial}_j\tilde{\chi}\tilde{\partial}^i\tilde{\partial}^j\tilde{\chi}\right],
\end{eqnarray}
where we have used Eq.~\eqref{eq:cvg_eom_0_eq1}, so that $H\dot{\varphi}=\dot{H}\varphi$, to eliminate $\dot{\varphi}$, and $'$ denotes the dimensionless derivative with respect to $\ln(a)$. If we define the following two dimensionless and time-dependent functions
\begin{equation}\label{eq:xxxxxx}
    \beta(a) \equiv \frac{\tilde{\beta}_2}{\tilde{\beta}_3\tilde{\varphi}^2}-6\frac{\tilde{\beta}_3}{\tilde{\beta}_2}\left(2-\frac{E'}{E}\right)E^2+\frac{\tilde{\beta}_3}{\tilde{\beta}_F}-\frac{1}{2}\tilde{\beta}_{3}\tilde{\varphi}^2,
\end{equation}
and
\begin{equation}\label{eq:rc_cvg}
    R_c(a) \equiv \frac{1}{\tilde{\varphi}},
\end{equation}
the equation can be further simplified to
\begin{equation}\label{eq:chi_eom_code_units}
     \tilde{\partial}^2\tilde{\chi} + \frac{R_c^2}{\beta a^4} \left[\left(\tilde{\partial}^2\tilde{\chi}\right)^2-\tilde{\partial}_i\tilde{\partial}_j\tilde{\chi}\tilde{\partial}^i\tilde{\partial}^j\tilde{\chi}\right] = \frac{3}{2\beta}\Omega_{m}a\left(\tilde{\rho}-1\right).
\end{equation}
This has a very similar form to the corresponding equations in the DGP or the cubic scalar Galileon model. Note that $|R_c|$ plays a similar role as the crossover radius in the DGP braneworld model.

One can again use Eq.~\eqref{eq:cvg_eom_0_eq1} to further simplify $\beta(a)$ by eliminating $\tilde{\varphi}$ as
\begin{equation}
    \beta(a) = -3\frac{\tilde{\beta}_3}{\tilde{\beta}_2}\left(1-2\frac{H'}{H}\right)E^2 + \tilde{\beta}_3 + 3\Omega_P\frac{\tilde{\beta}_3}{\tilde{\beta}_2}E^{-2},
\end{equation}
where we have used $\tilde{\beta}_F=1$ and Eq.~\eqref{eq:omega_cvg}. Using the following relations
\begin{eqnarray}\label{eq:tmp11}
    E^{-2} &=& \frac{1}{2\Omega_P}\left[\sqrt{\Omega_m^2a^{-6}+4\Omega_P}-\Omega_ma^{-3}\right],\\
    E^2-2EE' &=& 2\Omega_ma^{-3} + 2\frac{\Omega_m^2a^{-6}+\Omega_P}{\sqrt{\Omega_m^2a^{-6}+4\Omega_P}},
\end{eqnarray}
and using again Eq.~\eqref{eq:omega_cvg} this further becomes
\begin{equation}\label{eq:func_beta}
    \beta(a) = \frac{1}{2}\left(\frac{\tilde{\beta}_3}{2\Omega_P}\right)^{1/3}\left[5\Omega_ma^{-3}+\frac{3\Omega_m^2a^{-6}}{\sqrt{\Omega_m^2a^{-6}+4\Omega_P}}\right] + \tilde{\beta}_3.
\end{equation}
Similarly
\begin{equation}\label{eq:func_Rc2}
    R^2_c(a) = \frac{1}{2}\tilde{\beta}_3^{2/3}\left(2\Omega_P\right)^{-2/3}\left[\Omega_ma^{-3}+\sqrt{\Omega_m^2a^{-6}+4\Omega_P}\right].
\end{equation}

The top-right panel of Fig.~\ref{fig:background_factors} shows how $R^2_c(a)$ depends on the model parameter $\tilde{\beta}_3$.
Note that both functions, $\beta(a)$ and $R_c(a)$, are fully fixed by specifying $\Omega_m$ and $\tilde{\beta}_3$. Therefore there is one free parameter in this model, given by $\tilde{\beta}_3>0$.

\subsection{Vainshtein screening}
One of the key quantities in models employing the Vainshtein screening mechanism is the distance to the source, called the Vainshtein radius, $r_V$, where the linear perturbation analysis breaks down and the theory enters the non-linear regime. For scalar field models with derivative self-interactions it is the cubic- and higher-order terms that produce Vainshtein screening. It has been demonstrated that to explain the late-time cosmic acceleration, the Proca field has to be very light, $b_2=m^2 \to 0$, and a non-zero coupling coefficient $b_3$ activates the screening mechanism to ensure the theory is consistent with solar-system tests of gravity \cite{DeFelice:2016cri}.

We have seen in Eq.~\eqref{eq:chi_eom_code_units} that the non-linear term, which is what produces Vainsthein screening, is determined by $\beta(a)$ and $R_c^2(a)$, both of which depend on the free model parameter $b_3$, or its code-unit counterpart $\tilde{\beta}_3$. To make an educated choice of $\tilde{\beta}_3$, we compare the cvG model with the sDGP and csG models, whose behaviour has been well understood. To do this fairly, we follow \cite{Barreira:2013eea} (for the case of csG) to re-scale $\tilde{\chi}$ such that the source term of Eq.~\eqref{eq:chi_eom_code_units} becomes exactly identical to that of the EOM of the sDGP brane-bending mode as given by Eq.~(18) in \cite{Li:2013nua}; then we can simply compare the coefficients of the non-linear terms in these equations to decide for which values of $\tilde{\beta}_3$ does csG have a stronger Vainshtein screening than sDGP. This is achieved by introducing the redefined scalar mode, $\tilde{\chi}'$, as
\begin{equation}\label{field-perturbation-redefinition}
    \tilde{\chi} = \frac{3\beta_{\rm sDGP}}{2\beta}\tilde{\chi}',
\end{equation}
where we have used the $\beta_{\rm sDGP}$ function, which describes the coupling strength to matter of the brane-bending mode in the sDGP model given by
\begin{equation}\label{eq:beta_DGP}
    \beta_{\rm sDGP} = -\frac{\frac{1}{2}\Omega_{m}a^{-3} + \Omega_{rc}}{\sqrt{\Omega_{rc}\left(\Omega_{m}a^{-3} + \Omega_{rc} \right)}},
\end{equation}
with a typical value $\Omega_{rc} = \frac{1}{4H_0^2R^2_c} = 0.25$. In this case, Eq.~\eqref{eq:chi_eom_code_units} can be rewritten as the following equation for $\tilde{\chi}'$:
\begin{equation}\label{eq:rescaled_chi_eom_code_units}
    \tilde{\partial}^2\tilde{\chi}' + \frac{1}{3\gamma a^4} \left[\left(\tilde{\partial}^2\tilde{\chi}'\right)^2-\left(\tilde{\partial}_i\tilde{\partial}_j\tilde{\chi}'\right)\left(\tilde{\partial}^i\tilde{\partial}^j\tilde{\chi}'\right)\right] = \frac{1}{\beta_{\rm sDGP}}\Omega_{m}a\left(\tilde{\rho}-1\right),
\end{equation}
where the source term on the right-hand side is now identical to that in the sDGP equation \cite{Li:2013nua}, and we have defined a new time-dependent function
\begin{equation}
    \gamma(a) \equiv \frac{2\beta^2}{9\beta_{\rm sDGP}R^2_c}.
\end{equation}
Similarly, the Poisson equation, Eq.~\eqref{eq:poisson_code}, should be changed to
\begin{equation}\label{eq:rescaled_poisson_code}
    \tilde{\partial}^2\tilde{\Phi} = \frac{3}{2}\Omega_{m}a\left(\tilde{\rho}-1\right) + \frac{3\beta_{\rm sDGP}}{2\beta}\alpha\tilde{\partial}^2\tilde{\chi}'.
\end{equation}
From here on, without otherwise specified, we will drop the prime in $\tilde{\chi}'$ to lighten our notations.

To have a sense of the effect of Vainshtein mechanism analytically, we consider a static spherically symmetric top-hat density distribution of radius $\tilde{R}$ with the enclosed mass $\tilde{M}$ being
\begin{equation}
    \tilde{M}(\tilde{r}) \equiv 4\pi\int_0^{\tilde{r}}\left(\tilde{\rho}(\xi)-1\right)\xi^2d\xi,
\end{equation}
where we are using code units, such that $\tilde{\rho}$ is defined as in Eq.~\eqref{eq:code_unit}; $\tilde{r}$ is also in code unit such that $\tilde{r}=r/L$ and similarly $\tilde{R}=R/L$ with $R$ being the radius of the top-hat in physical unit. Note that $\tilde{\rho}=1$ outside the top-hat, so that the mass $\tilde{M}$ stops growing and becomes a constant at $\tilde{r}\geq\tilde{R}$.

We relate the mass distribution to $\tilde{\chi}$ using Eq.~\eqref{eq:rescaled_chi_eom_code_units}. Realising that $\tilde{\chi}$ depends only the radial coordinate, $\tilde{r}$, we obtain
\begin{equation}\label{eq:spherical_eq}
    \frac{1}{\tilde{r}^2} \frac{d}{d\tilde{r}} \left( \tilde{r}^2 \tilde{\chi} ,_{\tilde{r}} \right) + \frac{2}{3\gamma a^4} \frac{1}{\tilde{r}^2 } \frac{d}{d\tilde{r}} \left(\tilde{r}\tilde{\chi},_{\tilde{r}}^2\right) = \frac{\Omega_{m}a}{\beta_{\rm sDGP}}\left(\tilde{\rho}-1\right),
\end{equation}
where $,_{\tilde{r}}\equiv d/d\tilde{r}$. We integrate over the top-hat density distribution to yield
\begin{equation}\label{eq:spherical_eq_integrated}
    \tilde{\chi},_{\tilde{r}} + \frac{2}{3\gamma a^4}\frac{1}{\tilde{r}}\tilde{\chi},_{\tilde{r}}^2 = \frac{\Omega_{m}a}{4\pi\beta_{\rm sDGP}}\frac{\tilde{M}(\tilde{r})}{\tilde{r}^2}.
\end{equation}
Solving this second-order algebraic equation for $\tilde{\chi},_{\tilde{r}}$ we get
\begin{equation}\label{eq:algebraic_solution_in}
    \tilde{\chi},_{\tilde{r}} = \frac{4\tilde{R}^3}{3\beta_{\rm sDGP} \tilde{R}_V^3}\left[\sqrt{\left(\frac{\tilde{R}_V}{\tilde{R}}\right)^3 + 1} - 1\right]\tilde{F}_{{\rm N}}(\tilde{r}),
\end{equation}
for $\tilde{r}\leq\tilde{R}$, where we substituted the Newtonian acceleration in code unit (which can be solved using Eq.~\eqref{eq:rescaled_poisson_code} without taking into account the Proca field contributions),
\begin{equation}
   \tilde{F}_{{\rm N}}(\tilde{r}) = \frac{d\tilde{\Phi}}{d\tilde{r}} = \frac{3\Omega_ma}{8\pi}\frac{\tilde{M}(\tilde{r})}{\tilde{r}^2},
\end{equation}
and
\begin{equation}\label{eq:algebraic_solution_out}
    \tilde{\chi},_{\tilde{r}} = \frac{4\tilde{r}^3}{3\beta_{\text{sDGP}} \tilde{R}_V^3}\left[\sqrt{\left(\frac{\tilde{R}_V}{\tilde{r}}\right)^3 + 1} - 1\right]\tilde{F}_{{\rm N}}(\tilde{r}),
\end{equation}
for $\tilde{r}>\tilde{R}$, where the Newtonian acceleration in code unit becomes
\begin{equation}
   \tilde{F}_{{\rm N}}(\tilde{r}) = \frac{3\Omega_ma}{8\pi}\frac{\tilde{M}(\tilde{R})}{\tilde{r}^2}.
\end{equation} 
Here we identified the Vainshtein radius (in code unit) to be
\begin{equation}\label{eq:vainshtein_radius}
    \tilde{R}_V^3 = \frac{8\tilde{c}^2 \tilde{R}_S}{9\beta_{\text{sDGP}}\gamma a^3} = \frac{4\tilde{c}^2R_c^2\tilde{R}_S}{\beta^2a^3},
\end{equation}
where $\tilde{R}_S \equiv 3\Omega_m\tilde{M}(\tilde{R})/(4\pi\tilde{c}^2)$ is the Schwarzschild radius of the source in code unit\footnote{We note that the screening mechanism in the cubic-order Proca theory has been previously studied in Ref.~\cite{DeFelice:2016cri}. However, the equations for the temporal component, $\varphi$, are different in Ref.~\cite{DeFelice:2016cri} and this paper, probably because there $\varphi\equiv A^{0}$ and here we defined $\varphi\equiv A_{0}$. As a result, a direct comparison of the solutions of $\tilde{\chi},_{\tilde{r}}$ between these two papers is difficult and not pursued here. We have, however, checked that our spherical equation for $\varphi$ agrees with that given in Ref.~\cite{Heisenberg:2017hwb} in the weak-field limit.}. The physical meaning of this mechanism can be seen by considering the two limits, $\tilde{r},\tilde{R} \ll \tilde{R}_V$ and $\tilde{r} \gg \tilde{R}_V$. In the former case the solution Eq.~\eqref{eq:algebraic_solution_in} applies and we obtain, according to Eq.~\eqref{eq:rescaled_poisson_code}, the following result for the fifth-force (in code unit),
\begin{equation}
    \frac{3\beta_{\text{sDGP}}}{2\beta}\alpha\frac{d\tilde{\chi}}{d\tilde{r}} \to 2\frac{\alpha}{\beta} \left(\frac{\tilde{R}}{\tilde{R}_V}\right)^{\frac{3}{2}}\tilde{F}_{\rm N}(\tilde{r}) \ll \tilde{F}_{{\rm N}}(\tilde{r}),
\end{equation}
which represents the regime in which the fifth-force is strongly suppressed. In the latter case, we find
\begin{equation}
    \frac{3\beta_{\text{sDGP}}}{2\beta}\alpha\frac{d\tilde{\chi}}{d\tilde{r}} \to \frac{\alpha}{\beta}\tilde{F}_{{\rm N}}(\tilde{r}),
\end{equation}
which shows that the fifth-force takes a constant ratio $\alpha/\beta$ to the Newtonian acceleration.

\begin{figure}[h!]
    \centering 
    \includegraphics[width=.98\textwidth]{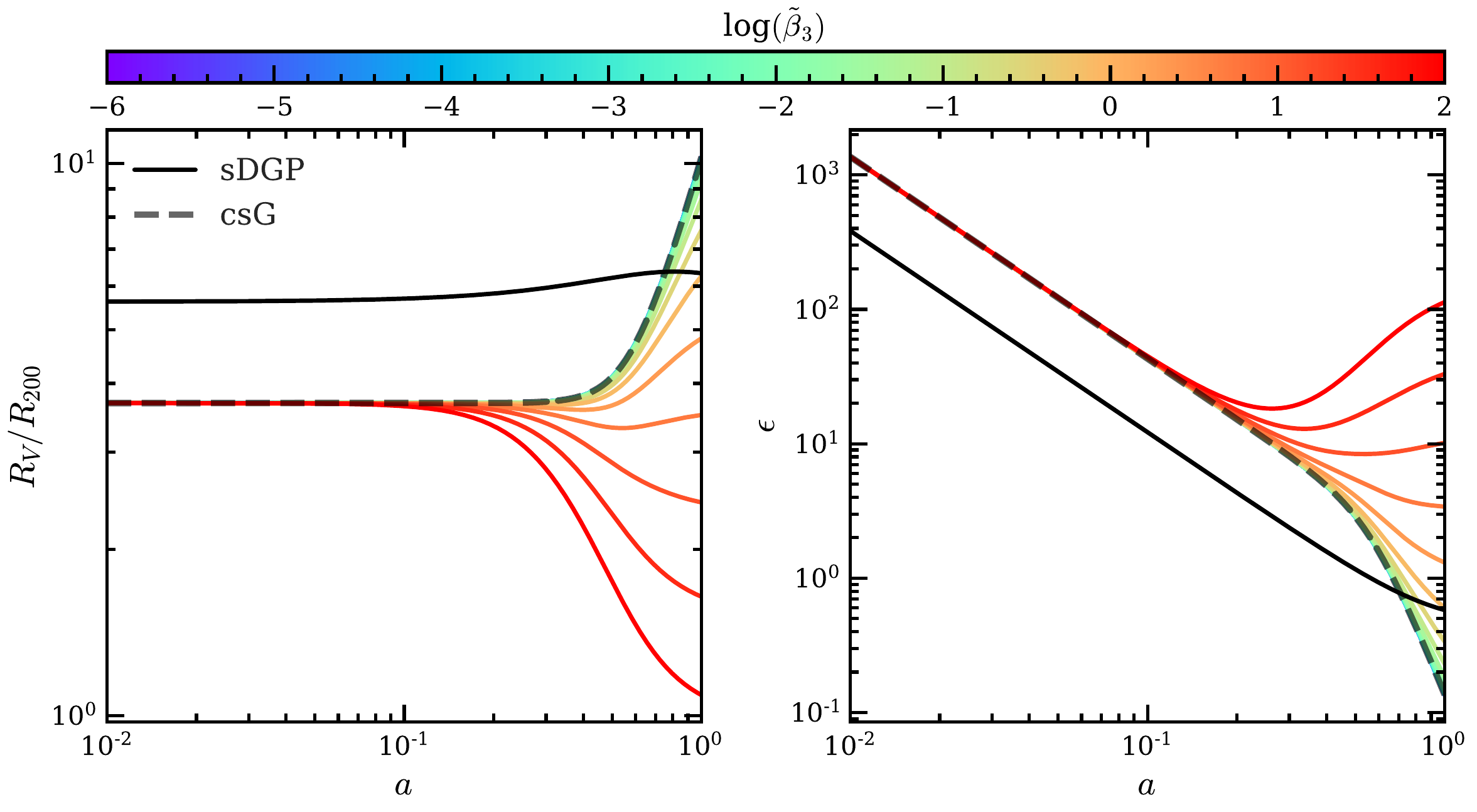}
    \caption{{\it left:} The time evolution of the relation between Vainshtein radius and top-hat radius for a given body. {\it right:} Coefficient of the non-linear derivative terms of the re-scaled scalar field equations. The figure compares the cvG model for different $\tilde{\beta}_3$ model parameters (colored solid lines) to sDGP (black solid lines), csG (grey dashed line).
    }
    \label{fig:fig_rv_nonlin}
\end{figure}

In the left panel of Fig.~\ref{fig:fig_rv_nonlin} we show the ratio between the Vainshtein radius $\tilde{R}_V$ and the top-hat radius $\tilde{R}_{200}$, for different values of $\tilde{\beta}_3$ (coloured solid lines), and compare to sDGP (black solid line) and csG (dashed line). Note, that due to the different background expansions this is not a fair comparison of sDGP to csG and cvG. In order to calculate the ratio, we have assumed that the spherical top-hat has a constant density within $\tilde{r}\leq\tilde{R}$ which is equal to $200$ times the critical density $\rho_c(a) = 3H(a)^2/8\pi G$ and equals
\begin{equation}\label{eq:r_200}
    R_{200}^3 = \frac{3 M_{200}}{4\pi 200 \rho_c(a)},
\end{equation}
making the ratio between $R_V$ and $R_{200}$ (note that here we ignore the tildes as this is equal to the ratio between the code-unit versions $\tilde{R}_V$ and $\tilde{R}_{200}$)
\begin{equation}\label{eq:r_200}
    \left(\frac{R_V}{R_{200}}\right)^3 = \frac{800R_c^2}{\beta^2} \left(\frac{H}{H_0}\right)^2 = \frac{800R_c^2}{\beta^2}E^2.
\end{equation}
For clarity we write down the corresponding equations for each of the considered cosmologies,
\begin{equation}\label{eq:rV_over_r200_comp}
    \frac{R_V}{R_{200}} = 
    \left \{
        \def\arraystretch{2.2}
          \begin{array}{lc}
            \sqrt[\leftroot{-1}\uproot{2}\scriptstyle 3]{\frac{1600 R^2_cE^2}{9\beta_{\rm sDGP}^2}}, & {\rm sDGP}\, , \\
            \sqrt[\leftroot{-1}\uproot{2}\scriptstyle 3]{\frac{1600E^2}{9\beta_{1,{\rm csG}}\beta_{2,{\rm csG}}}}, & {\rm csG}\, , \\
            \sqrt[\leftroot{-1}\uproot{2}\scriptstyle 3]{\frac{800 R^2_cE^2}{\beta^2}}, & {\rm cvG}\, ,
          \end{array}
       \right.
\end{equation}
where $\beta_{1,{\rm csG}}$ and $\beta_{2,{\rm csG}}$ are $\beta$ functions defined for the csG model in Ref.~\cite{Barreira:2013eea} (to avoid confusion with the $\beta$ function for the cvG model in this paper we have added a csG label to the subscript of its $\beta$'s, separated by a comma).

It can be seen from the left panel that the Vainshtein radius in the cvG model is insensitive to $\tilde{\beta}_3$ at early times, but becomes very strongly dependent on $\tilde{\beta}_3$ at $a\gtrsim0.1$. For example, choosing a $\tilde{\beta}_3 \sim \mathcal{O}(100)$ results in a screening radius that is nearly an order-of-magnitude smaller than its csG counterpart (the dashed line, which corresponds to $\tilde{\beta}_3 \to 0$) at $a\simeq1$, setting it approximately equal to the size $R_{200}$ of the over-density itself (note that at $a\simeq1$ we have $E\simeq1$).

In the right panel of the same figure we show the time evolution of the coefficient of the non-linear derivative terms of the re-scaled scalar field equations, as given in Eq.~\eqref{eq:rescaled_chi_eom_code_units} for the cvG model. This coefficient can be thought of as the controlling strength of the Vainshtein screening -- the larger it is, the more efficient the screening becomes. Because it is also present in the sDGP and csG cosmology, we show a comparison to the sDGP and the re-scaled csG model. Instead of showing the coefficients themselves, we have defined a new quantity $\epsilon$ as follows,
\begin{equation}\label{eq:fig_rv_nonlin_yaxis}
    \epsilon = 
    \left \{
        \def\arraystretch{2.2}
          \begin{array}{lc}
              -R_c^2 / \beta_{\text{sDGP}}, & {\rm sDGP}\, , \\
              -\beta_{1,{\rm csG}}\beta_{2,{\rm csG}}/ \beta_{\text{sDGP}}, & {\rm csG}\, , \\
              -\gamma, & {\rm cvG}\, .
          \end{array}
       \right.
\end{equation}
Again we note that values of $\tilde{\beta}_3 < \mathcal{O}(0.01)$ seem to closely mimic the csG model behaviour. While for $\tilde{\beta}_3\sim \mathcal{O}(100)$ there is less efficient screening, we can now see that the fifth-force starts to become weaker compared to the csG model starting from $z \approx 4$, ending with a $\epsilon$ that is $\sim\mathcal{O}(3)$ larger today.

The fact that $R_V/R_{200}$ and $\epsilon$ of the cvG model approach their corresponding values in the csG model for $\tilde{\beta}_3 \to 0$ deserves a couple of comments here. First, as mentioned earlier, the dynamics of the csG model depends on the initial condition of the scalar field, and different initial conditions can lead to different late-time behaviour. However, as we consider the tracker solution of the csG model, the late-time model behaviour show in Fig.~\ref{fig:fig_rv_nonlin} is a unique limiting case.

Second, it may seem that, because $\tilde{\beta}_3 \propto b_3$, as $\tilde{\beta}_3 \to 0$ we have $b_3 \to 0$, and we would expect the $G_3$ term in the Proca Lagrangian vanishes and the theory goes back to the GR limit with a massive vector field, rather than the csG limit. Here we distinguish between two scenarios. The first is to keep $\tilde{\beta}_2$ (or equivalently $b_2$) fixed while reducing $\tilde{\beta}_3$ (or $b_3$): here we do get back to the GR limit but the expansion history will also be dependent on $\tilde{\beta}_3$ -- this is {\it not} the scenario followed in this paper. 
The second scenario is to keep the background expansion history fixed and decrease $\tilde{\beta}_3$: then according to Eq.~\eqref{eq:omega_cvg} $\tilde{\beta}_2$ decreases accordingly; this is the scenario of this paper. 
In this case, there is a special scaling degeneracy which exists for Galileon-type models (see, e.g., Section IIIB of \cite{Barreira:2013jma} for a more detailed discussion), which we briefly review here. For simplicity, let us assume that the Proca vector field has only a longitudinal mode, and so the Lagrangians $\mathcal{L}_{2,3}$ can be schematically written as 
\begin{eqnarray}\label{eq:csg-lag}
    \mathcal{L}_2 &\propto& b_2\nabla^\mu\chi\nabla_\mu\chi,\nonumber\\
    \mathcal{L}_3 &\propto& b_3\nabla^\mu\chi\nabla_\mu\chi\Box\chi.
\end{eqnarray}
If we multiply $b_2$ by $T^2$, multiply $b_3$ by $T^3$ and divide $\chi$ by $T$, with $T$ being an arbitrary constant, then the physics is unaffected. Therefore, reducing $\tilde{\beta}_2$ and $\tilde{\beta}_3$ simultaneously with $\tilde{\beta}^3_2/\tilde{\beta}^2_3$ fixed would keep the physics unchanged by increasing $\tilde{\chi}$ accordingly. This is what happens in the csG model. In the cvG model, the presence of the $\mathcal{L}_F$ Lagrangian slightly complicates things, and breaks this scaling degeneracy, but the scaling degeneracy can be approximately restored with $\tilde{\beta}_3 \to 0$ (or $b_3 \to 0$). To see this, let us look at Eqs.~\eqref{eq:xxxxxx} - \eqref{eq:chi_eom_code_units} and consider the limit where $\tilde{\beta}_{2,3} \to 0$ simultaneously with $\tilde{\beta}^3_2/\tilde{\beta}^2_3$ fixed. To be concrete, we introduce the following scalings (with $T\ll1$):
\begin{eqnarray}
    \label{eq:scaling1}\tilde{\beta}_2 & \to & T^2\tilde{\beta}_2,\\
    \label{eq:scaling2}\tilde{\beta}_3 & \to & T^3\tilde{\beta}_3,\\
    \label{eq:scaling3}\tilde{\chi} & \to & T^{-1}\tilde{\chi},\\
    \label{eq:scaling4}\tilde{\varphi} & \to & T^{-1}\tilde{\varphi},\\
    \label{eq:scaling5}R_c & \to & TR_c,
\end{eqnarray}
in which Eq.~\eqref{eq:scaling4} is needed for the rescaled quantities to still satisfy Eq.~\eqref{eq:yy2}, and Eq.~\eqref{eq:scaling5} is because of Eq.~\eqref{eq:rc_cvg}. Then, of the 4 terms on the right-hand side of Eq.~\eqref{eq:xxxxxx}, all scale as $T$ apart from $\tilde{\beta}_3/\tilde{\beta}_F$ -- however, because $\tilde{\beta}_F=1$, we can see that with $T \to 0$ the term $\tilde{\beta}_3/\tilde{\beta}_F$ goes to zero more quickly than the other three terms and can therefore be neglected in this limit, and the function $\beta$ scales as $T$ approximately. Then all terms in Eq.~\eqref{eq:chi_eom_code_units} scale as $T^{-1}$, which means that the physics encoded in this equation is unaffected by the scaling, which is exactly the case for the csG discussed in \cite{Barreira:2013jma}. The observation that in this limit the cvG model behaves similarly to csG can be explained by the fact that the only term contributed by $\mathcal{L}_F$ and involving $\tilde{\beta}_F$ (i.e., the $\tilde{\beta}_3/\tilde{\beta}_F$ term in Eq.~\eqref{eq:xxxxxx}) -- which has no counterpart in the csG model -- has been neglected (as well as the similarity between $\mathcal{L}_{2,3}$ in the two models).

\subsection{Linear growth of the density field}
\label{subsect:linear_growth}

\begin{figure}[h!]
    \centering 
    \includegraphics[width=.98\textwidth]{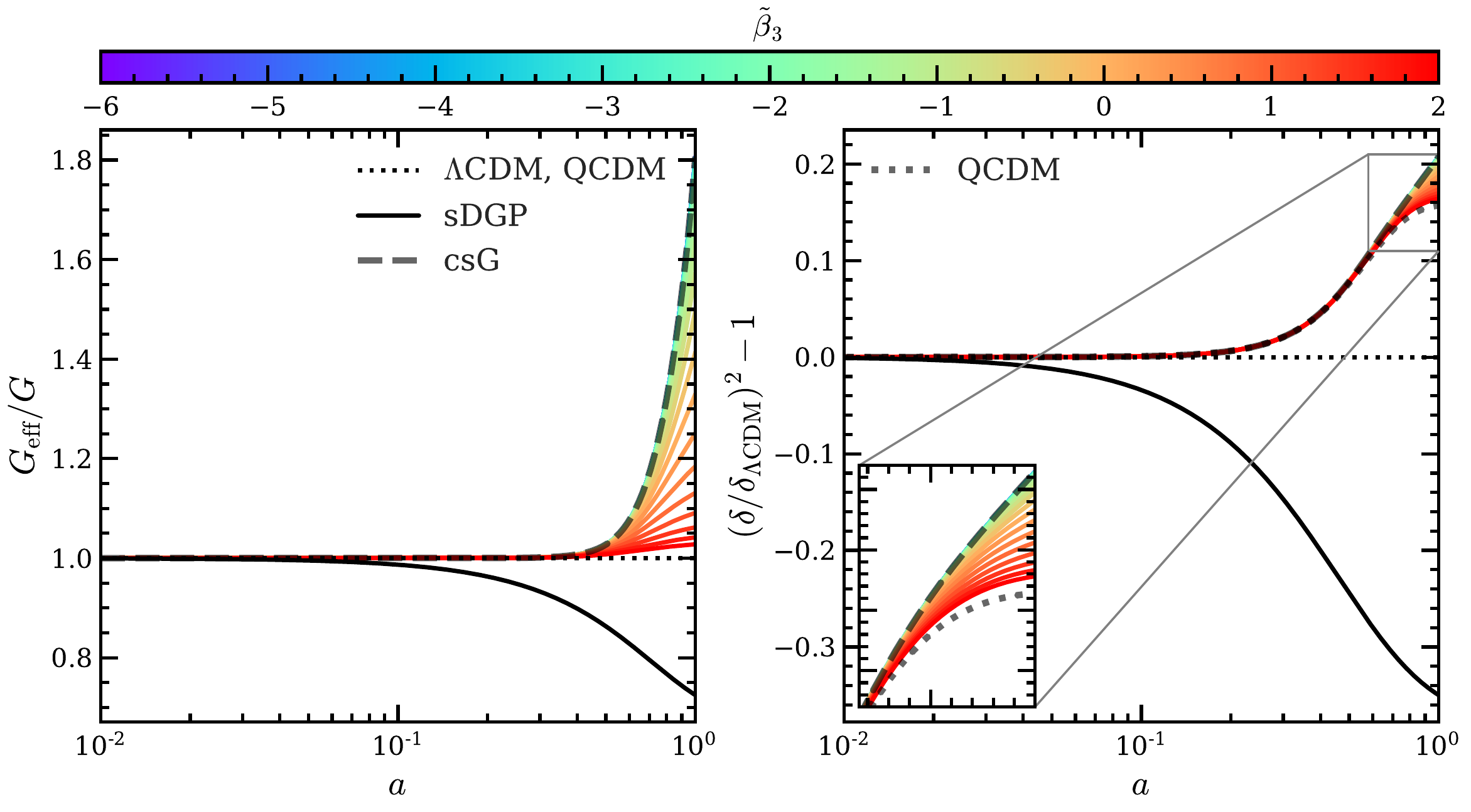}
    \caption{{\it Left:} Time evolution of the effective gravitational constant $G_{\rm eff}/G$. {\it Right:} Time evolution of the relative difference of the square of the density contrast. The figure compares the cvG model for different $\tilde{\beta}_3$ model parameters (colored solid lines) to sDGP (black solid lines), csG (grey dashed line), and $\lcdm$ (black dotted line).
    }
    \label{fig:fig_geff_D}
\end{figure}

Before we continue to explore late-time perturbations on sub-horizon scales, it is instructive to study the evolution of density fluctuations in linear perturbation theory. Of particular relevance is the linear rate of growth of cosmic structures, $\delta_{\rm M}(a) = D(a)\delta_0$, where $D$ is the normalized linear growth factor with $D(a=1)=1$. The growth is governed by
\begin{equation}\label{eq:linear_growth}
    D'' + \left(2+F\right)D' - \frac{3}{2}\frac{G_{\rm eff}}{G}\Omega_m(N)D = 0,
\end{equation}
where a prime denotes the derivative with respect to $N={\rm ln}(a)$ as before, $F = E'/E$ is the friction term, and $G_{\rm eff}/G$ is a time-dependent function that carries the modifications of the Newtonian potential, either due to a modified gravitational force or the clustering of dark energy. In the linear regime, each mode of the perturbed density field evolves independently. Their evolution is fully determined by $\Omega_m(N)$, $F$ and $G_{\rm eff}$. Note that in this paper we use $\Omega_m(N)$ to denote the matter density parameter at time $a$, to be distinguished from $\Omega_m$, which is the present-day value of the matter density parameter. To disentangle the relative importances of the modified gravitational strength $G_{\rm eff}$ and modified expansion history $E$ (or $F$) on the growth factor, we introduce the QCDM version of the cvG. The QCDM considers only modifications to the expansion history but not to the Newtonian potential, and is therefore identical for the cvG and csG models. For the set of considered models, the matter density parameter  evolves as
\begin{equation}\label{eq:time_dep_omega_m}
    \Omega_m(N) = \frac{\Omega_{m}e^{-3N}}{E^2},
\end{equation}
with $E^2$ given by Eq.~\eqref{eq:friedmann_code}. The friction coefficient for the different models can be written as,
\begin{equation}\label{eq:friction_terms}
    F = 
    \left \{
    \def\arraystretch{2.2}
    \begin{array}{ll}
        -\frac{3}{2}\Omega_{m}(N),  & \Lambda{\rm CDM}, \, \\
        -\frac{3}{2}\Omega_{m} \frac{e^{-3N}}{\sqrt{\Omega_{{rc}}}\sqrt{\Omega_{m}e^{-3N} + \Omega_{{rc}}} + \Omega_{m}e^{-3N} + \Omega_{{rc}}},  & {\rm sDGP}, \, \\
        \frac{1}{2} - \frac{1}{E^2}\left(\Omega_{m}e^{-3N} + \frac{\Omega^2_{m}e^{-6N} + (1-\Omega_{m})}{\sqrt{\Omega^2_{m}e^{-6N} + 4(1-\Omega_{m})}}\right), & {\rm cvG},\, {\rm csG},\, {\rm QCDM}. \,
    \end{array}
    \right.
\end{equation}

The modification of the Newtonian potential, which is proportional to the ratio between the fifth force, $F_5$, and the Newtonian gravity force, $F_{\rm N}$, is characterised by $G_{\rm eff}/G$ which in linear theory is given by the following time-dependent but scale-independent functions for the various models:
\begin{equation}\label{eq:g_effective}
    \frac{G_{\rm eff}}{G} = 1 + \frac{F_5}{F_{\rm N}} =
    \left \{
    \def\arraystretch{2.2}
    \begin{array}{ll}
        1,  & \Lambda{\rm CDM}, \, {\rm QCDM}, \, \\
        1 + \frac{1}{3 \beta_{\text{sDGP}}},  & {\rm sDGP}, \, \\
        1 - \frac{4 c_3\beta_{1,{\rm csG}}}{3\beta^2_{2,{\rm csG}}}, & {\rm csG}, \, \\
        1 + \frac{\alpha}{\beta}, & {\rm cvG}.\,
    \end{array}
    \right.
\end{equation}
The deviation of $G_{\rm eff}$ from $\lcdm$ for the various models can be seen in the left-hand panel of Fig.~\ref{fig:fig_geff_D}. To solve Eq.~\eqref{eq:linear_growth} we use the initial condition at $a_i = 0.01$: $D(a=a_i) = a_i$ and $D'(a=a_i)=1$, which correspond to the matter-dominated-era solution, $\delta \propto a$. The results can be seen on the right-hand panel of Fig.~\ref{fig:fig_geff_D}.

At early times, $a \lesssim 0.1$, $G_{\rm eff}/G \approx 1$ in all models, and therefore the differences from $\lcdm$ are mainly driven by the modified expansion history, $H$, and different matter densities $\Omega_m(a)$. In all modified gravity models except sDGP, both $H$ and $\Omega_m$ are larger than in $\lcdm$, so that their effects cancel out. The same happens in the sDGP cosmology though in this case $H$ and $\Omega_m$ are smaller than in $\lcdm$, and the growth of linear density perturbations is slightly slower. As a result, the relative difference $(\delta / \delta_{\lcdm})^2 - 1$ is almost zero in such early times.

At $a \gtrsim 0.1$, the evolution of $\delta$ is determined by the interplay of the modifications in $H$, $\Omega_m$, and $G_{\rm eff}/G$. We see how the modifications to effective gravitational constant enhance structure formation at late times for the cvG and csG models, while suppressing it in the sDGP model. As shown in the left panel of Fig.~\ref{fig:fig_geff_D}, for values of $\tilde{\beta}_3 \lesssim 0.01$, the evolution of $G_{\rm eff}$ in cvG is indistinguishable from that in csG. This, together with the fact that $H(z)$ and $\Omega_m$ are identical in the csG and cvG models, explains why in the right-hand panel of Fig.~\ref{fig:fig_geff_D} the evolutions of $\left(\delta/\delta_{\lcdm}\right)^2$ are also indistinguishable between csG and cvG with $\tilde{\beta}_3\lesssim0.01$. On the other hand, for large values of $\tilde{\beta}_3$, the behaviour of the cvG model approaches that of QCDM due to $G_{\rm eff}/G \to 1$\footnote{Note that it is possible to achieve a weaker gravity, $G_{\rm eff}/G < 1$, if one uses the full Lagrangian described in Eq.~\eqref{eq:proca_general}. With our restriction to the cubic order of the Lagrangian, we neglect the contributions of $\mathcal{L}_{4,5,6}$ which enter in very specific ways into $G_{\rm eff}$ as explained in the Ref.~\cite{DeFelice:2016uil}.}. This indicates that the cvG model, with a proper QCDM limit, could have a healthy behaviour regarding the ISW effect, which has proven to be an issue for the viability of the csG model. Cosmological constraints on the Proca theory have been studied in several works, e.g., Refs.~\cite{deFelice:2017paw,Nakamura:2018oyy,DeFelice:2020sdq} -- some of which actually have made use of the ISW data -- and these have placed strong constraints on the functional forms $G_2(X)$ and $G_3(X)$, disfavouring the simple model studied here with $G_2=G_3=X$. We will briefly comment on this and on the viability of the model in the end of Section \ref{subsect:beta_3_dependence}.

\section{Code tests and $N$-body Simulations}
\label{sec:simulations_section}

In this section we present the results of full $N$-body simulations based on the equations derived in the previous section. We begin in Section \ref{subsect:code_tests} with showing the outcomes of multiple tests which are essential for us to be confident about the reliability of the code. Afterwards in Section \ref{subsect:cosmological_simulation}, we present the results of the first set of the cosmological simulations of the simplified generalised Proca theory given in Section \ref{sec:proca_theory}. For details on the code algorithm we refer the reader to \cite{Li:2011vk} and \cite{Barreira:2013eea}.

\subsection{Code Tests}\label{subsect:code_tests}
All tests of the $N$-body code use a box-size of $L = 64$ Mpc$/h$, and a domain grid of $256^3$ cells with no grid refinement.

\subsubsection{One dimensional density fields}
The first set of tests is concerned with verifying the correct implementation of the linear terms in the cvG equation. By limiting ourselves to a one-dimensional matter distribution, the non-linear terms in the cvG equations simply vanish, and Eq.~\eqref{eq:rescaled_chi_eom_code_units} reduces to,
\begin{equation}
    \frac{d^2}{dx^2}\tilde{\chi}(x) = \frac{\Omega_{m}a}{\beta_{\text{sDGP}}}\delta(x).
\end{equation}
This means that an analytical expression can be easily obtained and comparable with the code results. Following \cite{Li:2011vk}, we first distribute the dark matter according to a one-dimensional sine field specified by,
\begin{equation}\label{eq:test_sine_dene}
    \delta(x) = -4\pi^2\frac{\beta_{\text{sDGP}}}{\Omega_m a} A {\rm cos}(2\pi x),
\end{equation}
such that the scalar field becomes
\begin{equation}
    \tilde{\chi}(x) = A {\rm cos}(2\pi x).
\end{equation}
We have performed the test with various values of $A$ and $\tilde{\beta}_3$. The result for $A = 10^{-8}$ and $\tilde{\beta}_3 = 10^{-6}$ can be seen in the left column of Fig.~\ref{fig:code_tests}, where the numerical solution (red dots), taken along a line which is parallel to the x-axis, are compared to the analytical solution (blue line) of $\tilde{\chi}$. In the top panel we show the chosen dark matter distribution, followed by the confirmation that the longitudinal mode, $\tilde{\chi}$, matches the analytical result. In the bottom panel we show just the x-component of the second partial derivative of the transverse mode, $\tilde{\partial}^2\tilde{B}_x$, as the y- and z-component share the same result. As the matter distribution is one dimensional, the source term in Eq.~\eqref{eq:B_eom_i} vanishes and therefore the transverse mode is expected to be zero. The fact that the numerical result for the transverse modes is zero, furthermore indicates not only that the linear terms are correctly implemented, but also that the non-linear source term of Eq.~\eqref{eq:B_eom_i} does not cause unwanted behaviour.

The second test uses a one dimensional Gaussian dark matter distribution, given by
\begin{equation}\label{eq:test_gauss_dene}
    \delta(x) = 1 + \frac{\beta_{\text{sDGP}}}{\Omega_m a}\frac{2\alpha}{\sigma^2}A\left(1 - 2\frac{(x-0.5)^2}{\sigma^2} \right)\text{exp}\left[-\frac{(x-0.5)^2}{\sigma^2}\right],
\end{equation}
and leads to a scalar field distribution of
\begin{equation}
    \chi(x) = A\left(1 - \alpha \text{exp}\left[-\frac{(x-0.5)^2}{\sigma^2}\right] \right).
\end{equation}
Again we have conducted multiple test for various values of $A$, $\sigma$, $\alpha$, and $\tilde{\beta}_3$. The result for $A = 10^{-6}$, $\sigma = 0.09$, $\alpha = 0.01$ and $\tilde{\beta}_3 = 10^{-6}$ can be seen in the central column of Fig.~\ref{fig:code_tests}, where the numerical (red dots) and analytical (blue line) are compared. Again, $\tilde{\chi}$ follows accurately the analytical result and the transverse mode vanishes with high precision.

\subsubsection{Three dimensional density fields}
\label{subsect:spherical-over-density-test}
After having performed tests for one dimensional matter distributions, we now move on to conduct more advanced tests using three dimensional distributions. This will reveal if there are any implementation errors of the non-linear terms, when they are needed. The simplest test in three-dimensions is the spherical symmetric top-hat distribution of matter. The analytical solution for $\tilde{r}\leq\tilde{R}$ is given by Eq.~\eqref{eq:algebraic_solution_in}, which can be re-written as
\begin{equation}\label{eq:tophat_solution_in}
    \frac{d\tilde{\chi}}{d\tilde{r}} = \frac{\beta^2 a^4}{6\beta_{\text{sDGP}} R^2_c}\left[\sqrt{\frac{4\Omega_m\delta_{\rm in} R^2_c}{\beta^2 a^3} + 1} - 1\right]\tilde{r},
\end{equation}
and Eq.~\eqref{eq:algebraic_solution_out} for $\tilde{r}>\tilde{R}$ which can be re-written as
\begin{equation}\label{eq:tophat_solution_out}
    \frac{d\tilde{\chi}}{d\tilde{r}} = \frac{\beta^2 a^4}{6\beta_{\text{sDGP}} R^2_c}\left[\sqrt{\frac{4\Omega_m\delta_{\rm out} R^2_c}{\beta^2 a^3}\left(\frac{\tilde{R}}{\tilde{r}}\right)^3 + 1} - 1\right]\tilde{r},
\end{equation}
where $\tilde{r}$ is the comoving coordinate scaled by the boxsize $L$, while $\tilde{R}$ is the radius of the spherical over-density scaled by $L$. The density inside the top-hat is given by $\delta_{\rm in}$ while it is $\delta_{\rm out}$ outside, which are both constants by definition.

\begin{figure}
    \centering 
    \includegraphics[width=.98\textwidth]{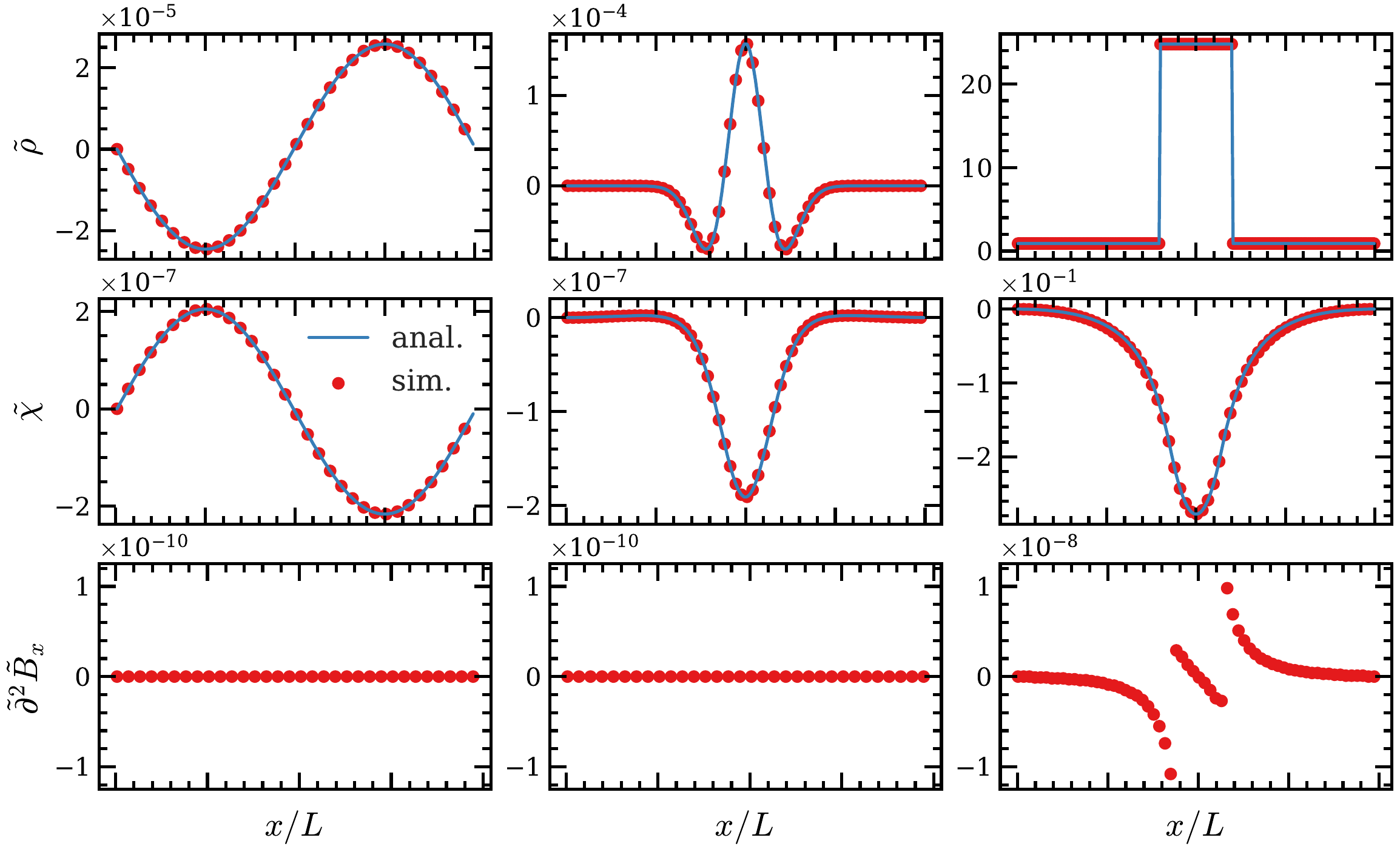}
    \caption{The various code tests conducted by assuming that the matter distribution is given by the following three ideal cases. {\it Left column:} A 1D sine-type matter density field as described by Eq.~\eqref{eq:test_sine_dene} with $A = 10^{-8}$. {\it Middle column:} A 1D Gaussian-shaped matter density field described by Eq.~\eqref{eq:test_gauss_dene}, with $A = 10^{-6}$, $\sigma = 0.09$, and $\alpha = 0.01$. {\it Right column:} A 3D spherical top-hat over-density with $\delta_{\rm in} = 23.77$, $\delta_{\rm out} = -0.1$, and $R_0 = 0.1$, as described in Section \ref{subsect:spherical-over-density-test}. For all three tests we have used $\tilde{\beta}_3=10^{-6}$, a simulation box of $L=64$ Mpc$/h$, and $256$ grid cells in each dimension. For each test and field quantity we compare the numerical result of the test simulations (red points), taken along a line which is parallel to the $x$-axis in a $y$-$z$ plane near the centre of the box, to its analytical solution (blue line).
    }
    \label{fig:code_tests}
\end{figure}

Given the value $\tilde{\chi}(\tilde{r} = 0)$, these equations can be integrated to find $\tilde{\chi}(\tilde{r} > 0)$ from its analytical expression. We call the $\tilde{\chi}(\tilde{r})$ obtained in this way the `analytical solution', even though in practice a numerical integration is required to get it. We tested various values of $\tilde{R}$, $\delta_{\rm in}$, and $\delta_{\rm out}$, where these values are always tuned in such a way as to make the average matter density $\tilde{\bar{\rho}} = 1$ (and equivalently the average $\delta=0$) in the entire simulation box. In the numerical implementation, the spherical top-hat is placed at the centre of the box, as illustrated in the upper right panel of Fig.~\ref{fig:code_tests}.

The middle and bottom panels of the right column of Fig.~\ref{fig:code_tests} shows the test result for a spherical top-hat of radius $\tilde{R} = 0.1$ with $\delta_{\rm in} = 23.77$ and $\delta_{\rm out} = -0.1$. We can see that the numerical result for $\tilde{\chi}$ (red points in the middle row), taken along a line which is parallel to the $x$-axis in a $y$-$z$ plane at the centre of the box, is in excellent agreement with the analytical solution (blue line), especially on small $\tilde{r}$. Far away from the centre, the agreement becomes less perfect since the analytical solution does not assume periodicity of the spherical density, while the numerical code uses periodic boundary condition so that the spherical density sees its own images.

With regards to $\tilde{\partial}^2\tilde{B}_x$, we can verify its accuracy by considering the analytical solution of $\tilde{B}_i$ in the spherical coordinate system centered on the top-hat. In this setting, the $\theta$- and $\phi$-components of $\tilde{B}_i$ vanish as $\tilde{\chi}$ only varies along the radial coordinate, $r$. Furthermore, as the transverse mode must obey the traceless condition $\nabla^i\tilde{B}_i=0$ and boundary condition $\tilde{B}_r(\tilde{r} = 0) = 0$, the radial component of $\tilde{B}_i$ has to vanish too. The numerical test solutions of $\partial^2B_i$, for $i=x$  and along the same axis as above, are shown as the red dots in the lower right panel of Fig.~\ref{fig:code_tests}, where we can see that it is indeed very close to zero, with a small nonzero amplitude of order $\mathcal{O}(10^{-8})$ due to numerical error and due to the fact that exact spherical symmetry is broken on a mesh of cubic cells.

\subsection{Cosmological simulations}
\label{subsect:cosmological_simulation}
Having verified the code implementation, we move on to run the code in a cosmological context with two objectives in mind. Firstly, we want to justify our assumptions, described in Section \ref{subsect:cosmo_eqns}, in which we neglect any 'backreaction' of $B_i$ on the evolution of $\chi$. Secondly, we want to study what influence the model parameter $\tilde{\beta}_3$ has on large-scale structure formation.

To this end, all simulations used in this section employ the same initial conditions, which were generated using {\tt 2LPTic} \cite{Crocce:2006ve}. The power spectrum of the initial density field,  at a scale factor of $a_{\rm ini} = 0.02$, assumes a flat $\lcdm$ cosmology obtained with {\tt CAMB} \cite{2011ascl.soft02026L}. One possible concern may be that, at this scale factor, differences of matter clustering are already present. However, judging from Fig.~\ref{fig:fig_geff_D}, at this time the difference between the growth factors of the cvG model with $\lcdm$ is well below sub-percent level. The fact that we use the same initial condition for simulations of different cosmologies ensures that the initial density fields have the same phases, and any differences at later times can solemnly be attributed the different dynamics and force laws. For comparisons, for every cvG simulation, we also run one for its QCDM counterpart, which has the expansion history of cvG but without modifications to the law of gravity.

The standard cosmological parameters used in the creation of the initial condition and simulations are
\begin{equation}\label{eq:cosmological_parameters}
    h = 0.6774, \quad \Omega_{\Lambda} = 0.6911, \quad \Omega_m=0.389, \quad \Omega_B=0.0223, \quad \sigma_8=0.8159
\end{equation}
(taken from the {\it Planck Collaboration} \cite{Ade:2015xua}). All cosmic simulations use a box-size of $L = 200$ Mpc$/h$, and a total number of dark matter particles of $N_p = 256^3$. The convergence criterion for the Gauss-Seidel algorithm is set to $|d^h| < \epsilon = 10^{-9}$. As it is not our objective to explore in great detail the predictions of various observables in the cvG model here, we use these small simulations in this paper to get a sense of the qualitative behaviours, and will report results from larger, higher-resolution simulations in follow-up works.

In $N$-body simulations for cubic and quartic scalar Galileon models, there is a well-documented problem that the numerical computation fails \cite{Barreira:2013eea,Li:2013tda} because the equation does not admit a physical solution under certain conditions \cite{Barreira:2013xea}. In the case of csG, this happens during a simulation when the scale factor $a\gtrsim0.8$ (the exact time at which this happens depends on the resolution, initial condition and cosmological parameters), in regions where matter density is very low, i.e., $\tilde{\rho} \to 0$. This problem can be traced to Eq.~\eqref{eq:spherical_eq_integrated}, which does not posses real solutions of $\tilde{\chi}_{,\tilde{r}}$ if
\begin{equation}
    \Delta \equiv 1 + \frac{12\Omega_m}{H_0^2a^3\tilde{r}^3}\int^{\tilde{r}}_0\left[\tilde{\rho}(\xi)-1\right]\xi^2d{\xi} < 0.
\end{equation}
There has been suggestion \cite{Winther:2015pta} that this is a real problem of the model itself, rather than a consequence of the approximations employed to simplify the field equations. Given that csG is a limiting case of the cvG model, we have found the same problem in our simulations for the latter, and followed the {\it ad hoc} fix employed in \cite{Barreira:2013eea} by setting $\Delta=0$ whenever the corresponding quantity becomes negative in a simulation mesh cell.

\subsubsection{The role of $B_i$ in cosmological simulations}
In order to confirm that the negligence of $B_i$ proposed in Section \ref{subsect:trans_eom} is justified, we ran a cosmological simulation with $\tilde{\beta}_3 = 10^{-6}$, and a domain grid of $256^3$ cells with no grid refinement.

\begin{figure}
    \centering 
    \includegraphics[width=.98\textwidth]{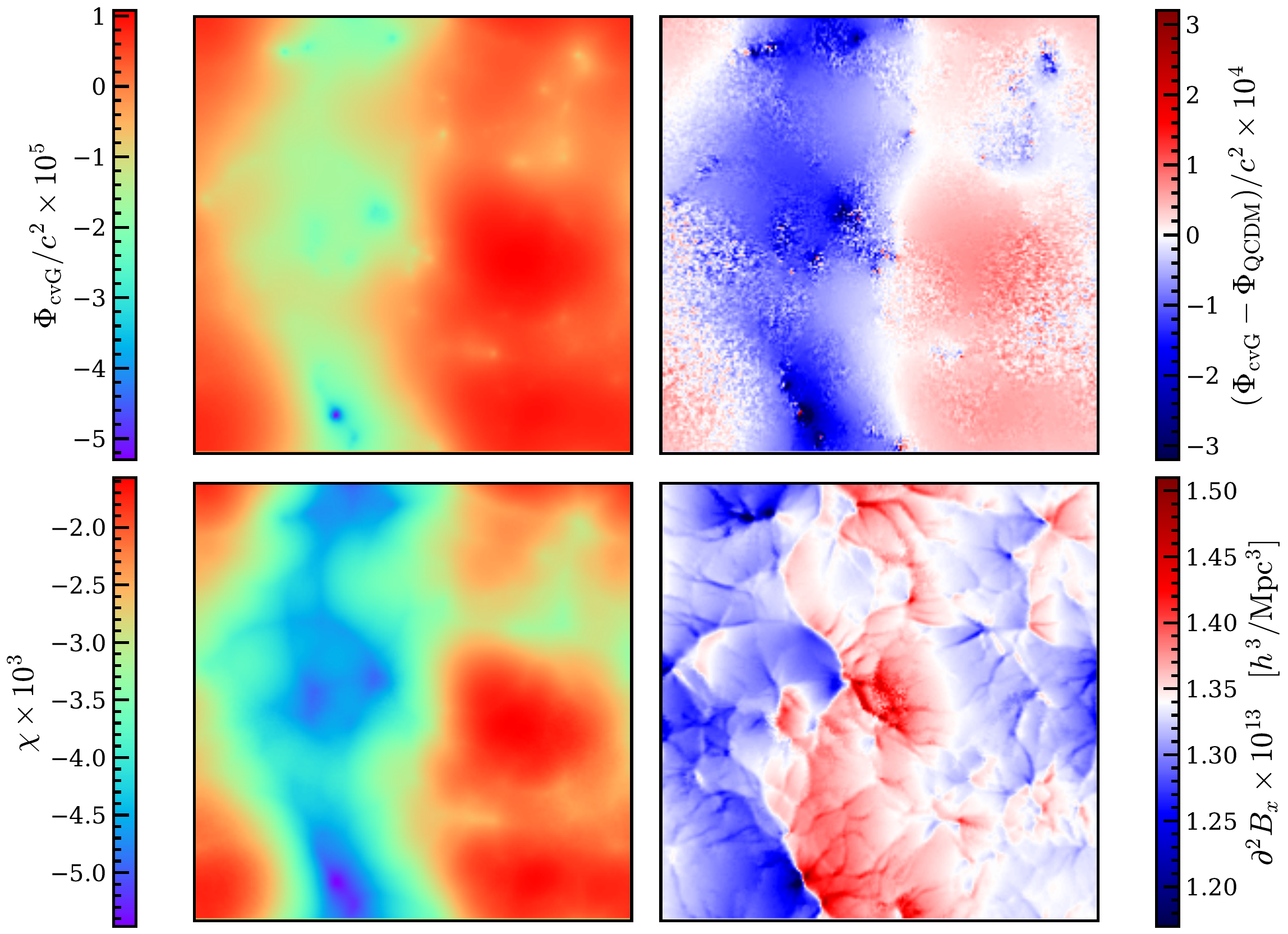}
    \caption{A visualisation of the spatial configurations of various fields taken from a slice of one cell size (with a thickness of $0.86$Mpc$/h$) in the simulations. {\it Top left:} Distribution of the total potential, $\Phi$, in the full cvG simulation. {\it Top right:} Difference between $\Phi$ in cvG and QCDM simulations, from which a stronger clustering in the former can be seen. {\it Bottom left:} The longitudinal vector mode, $\chi$. {\it Bottom right:} The second derivative of the transverse vector mode, $B_x$.
    }
    \label{fig:maps}
\end{figure}

A visualisation of the resulting fields including the gravitational potential and the extra degrees of freedom is shown in Fig.~\ref{fig:maps}. The maps have the same side length as the box, a depth of $0.86$ Mpc$/h$, and are cut out around the centre of the box. In the top row we show the gravitational potential $\Phi_{\rm cvG}$ (left) and the difference of $\Phi$ between cvG and its QCDM counterpart (right). As outlined earlier, the QCDM version only contains the background expansion and misses the fifth-force term which results in a weaker clustering of matter as compared to cvG. This is clearly visible in the right panel, where the blue (red) indicates a higher matter density around haloes 
in the cvG model (voids in the QCDM model).

In the bottom panels we present visualisations of the $\chi$ field (left) and the $\partial^2B_x$ component of the transverse mode (right) for the same slice of the simulation box.
The $\chi$ field, like the potential $\Phi$, is very smooth with a similar dependence on the underlying dark matter density and reaches local minima within halos and local maxima in voids. This is as expected as, apart from strongly screened regions, the fifth-force due to $\vec{\nabla}\chi$ generally has the same direction as and is proportional in magnitude to standard gravity. The distribution of $\partial^2B_i$ on the other hand is very rich in texture. This is because $\partial^2B_i$ is sourced by higher-order derivatives of $\chi$, cf.~Eq.~\eqref{eq:B_eom_i}. While the complexity of Eq.~\eqref{eq:B_eom_i} makes it difficult to interpret this map intuitively, we observe that it follows the patterns of the other maps in general.

\begin{figure}
    \centering 
    \includegraphics[width=.98\textwidth]{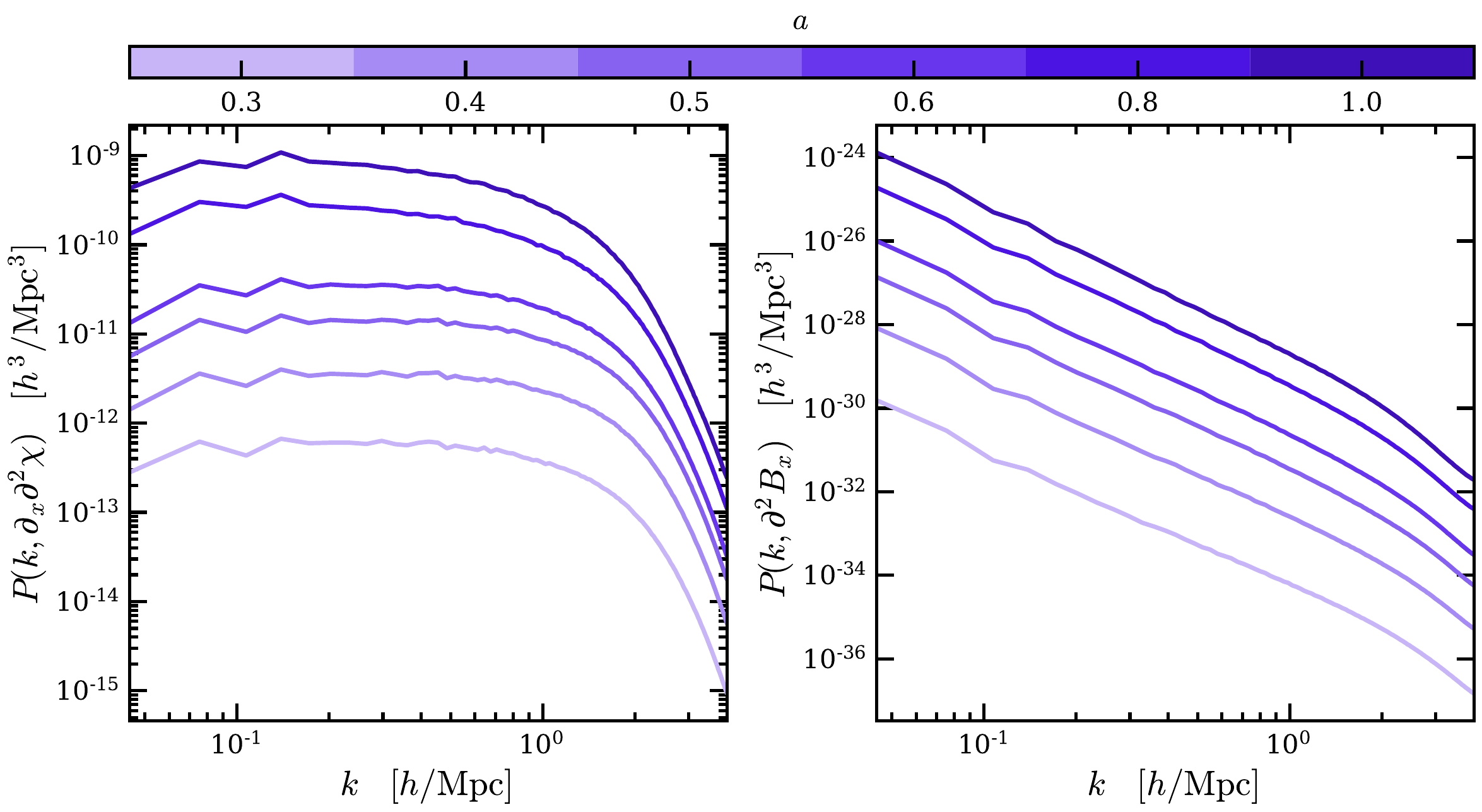}
    \caption{The power spectrum of (spatial derivatives of) the longitudinal (left) and transverse (right) mode of the Proca field, for $\tilde{\beta}_3 = 10^{-6}$. The different lines are results at different values of the scale factor $a$, as indicated by the colour bar on the top. Note the large amplitude differences between the two panels.}
    \label{fig:extra_dof_power_spectrum}
\end{figure}

While Fig.~\ref{fig:maps} intuitively shows the spatial configurations of various physical quantities in their own physical units, the comparison between the amplitudes of $\chi$ and $\partial^2B_i$ should not be used as a direct indicator to assess the relative importance of the longitudinal and transverse modes in affecting structure formation. Actually, from the field decomposition, Eq.~\eqref{eq:helmholtz} in Section \ref{sec:proca_theory}, we can see that a fairer comparison can be done by comparing the magnitudes of $\partial_i\chi$ and $B_i$. For simpler computation, we show the power spectra of $\partial_x\partial^2\chi$ and $\partial^2B_x$ at various times in Fig.~\ref{fig:extra_dof_power_spectrum}. Note that, because $\partial_i\partial^2\chi$ and $\partial^2B_i$ both have unit of $(h/{\rm Mpc})^{3}$, their power spectra have the unit of $(h/{\rm Mpc})^{3}$.

As the magnitude of the cvG longitudinal mode $\chi$ increases with matter density perturbations, the $P(k)$ of $\partial^2\partial_i\chi$, which we visualise for $a \in [0.3, 1.0]$ in Fig.~\ref{fig:extra_dof_power_spectrum}, also increases continuously as expected. It is interesting to note that while the matter power spectrum peaks $k\sim \mathcal{O}\left(10^{-2}\right)h/$Mpc, the power spectrum for $\partial^2\partial_i\chi$ has a significantly more flattened shape until $k\sim\mathcal{O}(1)h/$Mpc, which is because of the additional spatial derivative in $\partial_x\partial^2\chi$ (on large scales the power spectra of $\partial^2\chi$, $\partial^2\Phi$ and matter density are expected to have similar shapes because of the weak screening).

The right panel of Fig.~\ref{fig:extra_dof_power_spectrum} shows the time evolution of the power spectrum of $\partial^2B_x$. While this quantity also increases over time, we note that its amplitude is $\sim15$-$20$ orders of magnitude smaller than the $\partial_x\partial^2\chi$ power spectrum on all scales probed by the simulation. This serves as a confirmation that the transverse mode plays a very minor role compared with the longitudinal mode, on linear scales (as it was previously shown by \cite{DeFelice:2016uil}) as well as on non-linear scales. In particular, it verifies that it is a good approximation to neglect the terms involving $B_i$ in the vector field equation of motion \eqref{eq:cvg_eom_i}. This is the approximation that we shall take in what follows, and in future simulations of the cvG model.

\subsubsection{The dependence on $\tilde{\beta}_3$}
\label{subsect:beta_3_dependence}
We have seen above that, unlike the csG model, the cvG model has a free parameter which we choose to be represented by $\tilde{\beta}_3$. This parameter does not affect the background expansion history of the model, but controls the strength of the fifth-force of $G_{\rm eff}/G$, cf.~Eq.~\eqref{eq:g_effective} and Fig.~\ref{fig:fig_geff_D}. Also, in Fig.~\ref{fig:fig_rv_nonlin} we have seen that the degree of non-linear Vainshtein screening depends on $\tilde{\beta}_3$. As the screening effect on large-scale structure formation is most accurately captured by $N$-body simulations, here we give a first idea about this effect, while leaving a more detailed study of various non-linear observables in the cvG model to future works.

For this, we have run three cosmological simulations employing the full set of equations derived in Section \ref{sec:n_body_eqns} using $\tilde{\beta}_3 = (10^{-6}, 1, 100)$, using a domain grid of $256^3$ cells. The cells are refined when the effective number of particles $N_p > 9.0$ up until a finest resolution of $2^{16}$ cells per dimension (if they were to cover the whole simulation box) is reached. The simulations each ran in only about $1500$ core-hours, underlining the viability of much larger and better resolution simulations simulations with our code.

To get an understanding of the impact of $\tilde{\beta}_3$ on the cvG cosmology through a enhanced effective gravitational constant, $G_{\rm eff}/G$, and the Vainshtein screening, $R_V$, we have run four additional simulations using the same settings as outline above. One of these is the above-mentioned QCDM variant, which differs from a $\lcdm$ simulation only by a modified (cvG) background expansion history\footnote{Note that that the background expansion history is independent of $\tilde{\beta}_3$, so that only one QCDM simulation is needed.}, and is used to isolate the effect of the latter. For the other three sets of simulations, we neglect the non-linear terms in the EoM of $\chi$, which is equivalent to removing the screening mechanism by simply re-writing Eq.~\eqref{eq:rescaled_poisson_code} as,
\begin{equation}\label{eq:linearized_rescaled_poisson_code}
    \tilde{\partial}^2\tilde{\Phi} = \frac{3}{2}\Omega_{m}a\left(1 + \frac{\alpha}{\beta} \right)\left(\tilde{\rho}-1\right),
\end{equation}
using Eq.~\eqref{eq:g_effective}. These are what we call {\it linearised} simulations, and the comparison of them with the full simulations can illustrate the quantitative impact of the Vainshtein screening.

\begin{figure}[h!]
    \centering 
    \includegraphics[width=.98\textwidth]{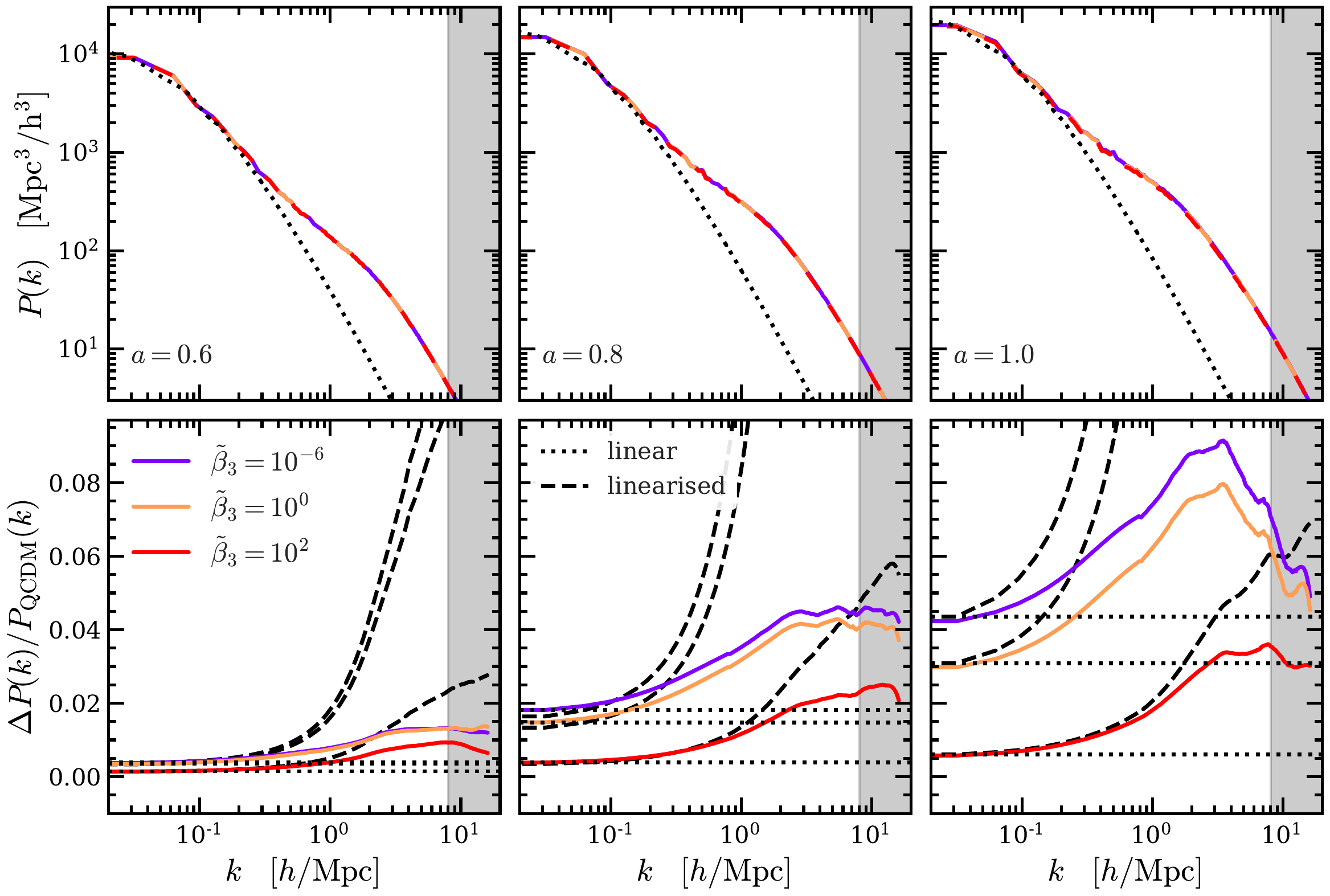}
    \caption{The matter power spectrum in the cvG model. Each column shows the results for a different scale factor: {\it left:} $a=0.6$, {\it centre:} $a=0.8$, {\it right:} $a=1.0$. {\it Top:} The matter power spectrum of linear perturbation theory (dotted) and the cvG model for three values of $\tilde{\beta}_3 = (10^{-6}, 1, 100)$, indicated by a purple, orange, and red line respectively. {\it Bottom:} Relative difference of the matter power spectra of the cvG and QCDM models, $\Delta P(k) / P_{\rm QCDM}(k) \equiv (P_{\rm cvG}(k) - P_{\rm QCDM}(k))/P_{\rm QCDM}(k)$. A Savitzky–Golay filter has been used to smooth $\Delta P(k) / P_{\rm QCDM}(k)$. Each panel compares linear perturbation theory (black dotted), to results obtained from full (coloured solid) and linearised (black dashed) simulations. The vertical grey shaded region in each panel indicates where $k>k_{\rm Ny}$ where $k_{\rm Ny}$ is the Nyquist frequency.
    }
    \label{fig:power_spectrum}
\end{figure}

Fig.~\ref{fig:power_spectrum} compares the linear matter power spectrum (black dotted lines) with the predictions by the linearised (black dashed) and fully non-linear (coloured) simulations, at $a=0.6$ (left), $a=0.8$ (centre) and $a=1.0$ (right). The linear power spectrum $P(k;z)$ is obtained by multiplying the initial power spectrum $P(k;z_{\rm ini})$ with $\left[D(z)/D(z_{\rm ini})\right]^2$, where $D$ is the linear growth factor discussed in Section \ref{subsect:linear_growth}. The non-linear matter power spectra are measured from the simulations using {\tt POWMES} \cite{Colombi:2008dw}. The relative difference of the matter power spectra of the cvG and QCDM models, $\Delta P(k) / P_{\rm QCDM}(k)$, has been smoothed using a Savitzky–Golay filter of third order with a kernel width of $51$ data-points. The shaded region in each panel indicates the regime of $k$ beyond the Nyquist frequency\footnote{Note that the Nyquist frequency, $k_{\rm Ny}$, marks the absolute maximum up to which we can the power spectrum can be trusted. First alterations can already appear at $k_{\rm Ny}/8$.}. The lower row of Fig.~\ref{fig:power_spectrum} shows the relative differences of the matter power spectra given by linear theory (dotted lines), linearised simulations (dashed) and full simulations (solid) with respect to their QCDM counterparts (i.e., QCDM linear theory and simulation predictions).

Fig.~\ref{fig:power_spectrum} allows for a number of conclusions. Firstly, we have seen in Section \ref{sec:n_body_eqns} that the csG model is a limiting case of the cvG model with $\tilde{\beta}_3\to0$, and the result in Fig.~\ref{fig:power_spectrum} confirms that the power spectrum in the case of $\tilde{\beta}_3=10^{-6}$ behaves similarly to what was found in  Ref.~\cite{Barreira:2013eea} for the csG model -- this serves as an independent check of the new numerical implementation in {\tt ECOSMOG}.

Secondly, as expected from Fig.~\ref{fig:fig_geff_D}, a larger value of $\tilde{\beta}_3$ leads to a smaller enhancement of matter clustering with respect to QCDM. We can also assess how effective the Vainshtein screening is for the different values of $\tilde{\beta}_3$ by comparing the results of the full (coloured solid lines) and linarised (black dashed) simulations in the bottom row. It becomes strikingly clear how the neglect of the non-linear terms in the EOM of $\chi$ leaves over-densities unscreened, leading to a much higher clustering power at small scales. The effect of the neglected screening mechanism propagates to larger scales the smaller $\tilde{\beta}_3$ is: at $a = 1$, scales of $k \gtrsim  4$ $h/$Mpc are screened for $\tilde{\beta}_3 = 10^{-6}$ and $1$, while for $\tilde{\beta}_3 = 100$ the clustering is only weakly damped. This is as expected from the left panel of Fig.~\ref{fig:fig_rv_nonlin}, which shows that the screening radius decreases when $\tilde{\beta}_3$ increases, meaning that for large values of $\tilde{\beta}_3$ the non-linear screening effect will be restricted to smaller scales and will be weaker. The observable peaks in the coloured lines in the lower panels, that becomes more pronounced with time, are a clear signature of the Vainshtein mechanism at work to bring gravity back to Newtonian on small scales. Interestingly, a qualitatively similar result has been obtained in Ref.~\cite{Heisenberg:2019ekf} based on the kinetic field theory.
 
Thirdly, we note that on large scales ($k<k_\ast$) the predictions by linear theory, the full and the linearised simulations all agree. The exact value of $k_\ast$ depends on redshift and the model parameter $\tilde{\beta}_3$. As an example, at $a=0.6$ we have $k_\ast\simeq0.3h/$Mpc for $\tilde{\beta}_3=100$ while $k_\ast\simeq0.15h/$Mpc for $\tilde{\beta}_3\leq1$; by the time $a=1$, however, $k_\ast$ has become much smaller for all $\tilde{\beta}_3$ values. The dependence on $\tilde{\beta}_3$ is due to the same reason as mentioned above, namely a larger $\tilde{\beta}_3$ means a smaller Vainshtein radius. The dependence on redshift is a combined consequence of the time evolution of the Vainshtein radius (cf.~the left panel of Fig.~\ref{fig:fig_rv_nonlin}) and the progressively non-linear matter clustering. Overall, the full simulation result actually agrees better with linear perturbation theory than the linearised simulation, due to the stronger fifth-force effect of the latter, and we can conclude that the screening mechanism does not affect the large scales typically associated with linear perturbation theory ($k\lesssim0.1h$/Mpc), which is therefore still a valid approximation on those scales.

Finally, we stress again that in this plot the linear matter power spectrum is calculated by simply extrapolating the initial power spectrum using the linear growth factor in the cvG model, rather than based on a full perturbation analysis. Therefore the good agreement between the linear theory and full simulation predictions can not be used as an evidence of the validity of the quasi-static approximation employed in this paper. However, it was shown, by using a modified version of {\tt CAMB}, in \cite{Barreira:2013eea} that the QSA does not have appreciable impact on $P(k)$ at large scales for the csG model which is very similar to the cvG model with $\tilde{\beta}_3=10^{-6}$. Comparing the behaviour of the relative difference between the full theory cvG and QCDM power spectra to the results of \cite{Barreira:2013eea} adds confidence on the applicability of the QSA on large scales for the cvG model.

Before finishing this subsection, let us briefly comment on the implication of the $\tilde{\beta}_3$-dependence of the fifth-force effect in the cvG model on its viability. The cvG model has an identical background expansion history to the csG model with the same cosmological parameters, and both do not have a $\lcdm$ limit, which suggests that the simple model studied in this paper could struggle in matching observations such as the CMB shift parameter and BAO \cite{deFelice:2017paw,Nakamura:2018oyy,DeFelice:2020sdq}. In Ref.~\cite{Barreira:2014ija}, including massive neutrinos was proposed as an alternative way to generalising $G_2$ and $G_3$ into non-linear functions of $X$ to bring compatibility of the csG model with those observations. But the simplest csG model with linear $G_2$ and $G_3$ still faces other challenges including void lensing (e.g., \cite{Barreira:2015vra,Baker:2018mnu}) and a wrong sign of the ISW effect (e.g., Refs.~\cite{Barreira:2014ija,Barreira:2014jha,Renk:2017rzu,Peirone:2017vcq}), due to a very fast increase of $G_{\rm eff}/G$ at late times (cf.~Fig.~\ref{fig:fig_geff_D}, $\tilde{\beta}_3=10^{-6}$). The cvG model with $\tilde{\beta}_3\gg1$ offers a potential way around this problem while maintaining other properties of the csG model, because $G_{\rm eff}/G$ can be strongly suppressed towards unity. We hope to revisit the cosmological constraints on the cvG model in a future publication.

\section{Discussion and conclusions}
\label{sect:discussion}
To summarise, in this paper we have performed the first self-consistent non-linear cosmological simulations of the generalized Proca theory, or the vector Galileon model, up to cubic order (cvG). This was achieved by adapting the \ecosmog simulation code, to implement the relevant equations under the weak-field and quasi-static approximations. We find that the cvG equation for the longitudinal mode of the vector field has the same structure in terms of spatial derivatives as, while differing in the time evolution of the coefficients from, the cubic scalar Galileon (csG) and sDGP models (see Section \ref{sec:proca_theory} for equations in natural units and Section \ref{sec:n_body_eqns} for equations in code units). In particular, cvG has the same background expansion history as csG for the same cosmological parameters. However, unlike the csG model, the cvG model has a free parameter $\tilde{\beta}_3$ which controls the strength of the fifth-force and the effectiveness of the Vainshtein screening.

We investigated the time evolution of various quantities including the screening radius, $R_V$, the magnitude of non-linear screening terms, $\epsilon$, the effective gravitational constant, $G_{\rm eff}$, and the relative difference of the density contrast $\delta/\delta_{\lcdm}$ (cf.~Fig.~\ref{fig:fig_rv_nonlin}, \ref{fig:fig_geff_D}). For all quantities we found their evolution in the cvG and csG models to be indistinguishable at early times, $a \lesssim 0.1$. This trend is continued at late times, $a \gtrsim 0.1$, for a cvG model parameter $\tilde{\beta}_3 \to 0$. If however $\tilde{\beta}_3 \to \infty$, than $R_V \to 0$, $\epsilon \to \infty$, $G_{\rm eff}/G \to 1$, and $\delta/\delta_{\lcdm}$ converges to the QCDM variant of the cvG model. This makes the cvG model more versatile and endows it with richer phenomenology.

In deriving the equations for $N$-body implementation, we have made a couple of simplifications. The first is that we have used the perturbed constraint equation satisfied by the temporal component of the Proca field, $\varphi$, to eliminate the time derivatives within the equation of motion for the longitudinal mode $\chi$, cf.~Eq.~\eqref{eq:chi_eom_i1}. This is done exactly, {\it without} resorting to the usual quasi-static approximation. The second is that we have manipulated the equation of motion for the transverse model, Eq.~\eqref{eq:B_eom_i_complex}, to obtain a much simplified approximate version, Eq.~\eqref{eq:B_eom_i}. This allows the $B^i$ field to be calculated easily in simulations, and allows the validity of the approximations used to be tested {\it a posteriori}.

We ran a set of moderate cvG cosmological simulations to investigate three questions. Firstly, proof that the transverse mode, $B_i$, is negligible compared with the longitudinal mode, $\chi$. By measuring their power spectra, we show that $P(k; \partial_i\partial^2 \chi)$ is about $15$-$20$ orders of magnitude larger than $P(k; \partial^2 B_i)$ on all scales probed by the simulation. Consequently, we expect the `back-reaction' of $B_i$ on the evolution of $\chi$ to be very small, justifying the neglect of the $B^i$ field in future simulations and confirming the findings of \cite{DeFelice:2016uil}.

Secondly, verification of the suppression of the fifth-force by the Vainshtein mechanism for the cvG model. To this end we have run cosmological simulations of the full cvG model and its {\it linearised} counterpart with $\beta_3 = (10^{-6}, 10^{0}, 10^{2})$. By comparing their relative power spectrum enhancement with respect to QCDM, $\Delta P(k)/P_{\rm QCDM}(k)$, the suppression of the fifth-force is quantified, c.f.~Fig.~\ref{fig:power_spectrum}. The comparison has made it clear how the neglect of the non-linear terms in the EOM of $\chi$ leaves over-densities unscreened, leading to a much higher clustering power at small scales. 

Finally, we show how the cvG model parameter $\tilde{\beta}_3$ affects the screening behaviour. The results in Fig.~\ref{fig:power_spectrum} confirm that the matter power spectrum in the case of $\tilde{\beta}_3=10^{-6}$ behaves similarly to what was found in  Ref.~\cite{Barreira:2013eea} for the csG model. However, the larger $\tilde{\beta}_3$ is, the smaller the enhancement of matter clustering with respect to QCDM becomes. The effect of the weakened screening mechanism also propagates to larger scales the smaller $\tilde{\beta}_3$ is: at $a = 1$, scales of $k \gtrsim  4$ $h/$Mpc are screened for $\tilde{\beta}_3 = (10^{-6}$, $1$), while for $\tilde{\beta}_3 = 100$ the clustering is only weakly damped. This agrees qualitatively with what we find in Fig.~\ref{fig:fig_rv_nonlin}, but the full non-linear simulations allow the effects to be more accurately quantified.

A more comprehensive investigation of the predictions of various physical quantities by the cvG model is need for better understanding the cosmological behaviours and observational implications of the model. This, however, requires more independent realisations of higher-resolution simulations covering more values of $\tilde{\beta}_3$, which are beyond the scope of this work and will be left for future work. We also note that, while this publication has focused on the simplest Proca theory at cubic order, with $G_2=G_3=X$, it should be very straightforward to extend our code to simulate models with generic non-linear functions for $G_{2,3}$ in the future. Such functions add further flexibilities to the generalised Proca theory -- indeed, ongoing research conducted by one of the authors has found that the GP theory up to cubic order can offer a better fit to available observational data than the standard $\lcdm$ model; see also Ref.~\cite{DeFelice:2020sdq} for some recent progress in developing linear Boltzmann codes for the GP theory.

Even for the simplest case with $G_2=G_3=X$, the cvG model's dependency on $\tilde{\beta}_3$ makes it an extension of the csG model from a phenomenological point of view, and this opens up possibilities to overcome the challenges the csG model faces in terms of void lensing and the ISW effect. These challenges originate from the fact that, if the csG field is the driving force of the accelerated cosmic expansion at late times, a byproduct is the quickly-deepening gravitational potential during this period. For the ISW effect, this is in contrast to $\lcdm$, where the potential becomes shallower due to the accelerated expansion, and therefore leads to a wrong sign of the ISW-galaxy correlation. As the deepening of the gravitational potential at late times can be weakened using an increased $\tilde{\beta}_3$, the cvG model offers a potential way around these issues, while maintaining the other properties of csG. We will investigate these possibilities in the future.

Finally, even though we have justified the neglect of the transverse mode of the vector field $B^i$ in cosmological simulations, it is possible that in other situations this is no longer a good approximation. For example, the Proca field does not have to be the driving force behind the accelerated cosmological background expansion, but might have effects on galactic scales and the transverse modes could give rise to a change of structure formation on such scales. With some appropriate adaption and extension, our code will be able to be used as a tool for investigations in such circumstances.

\acknowledgments

CB acknowledges support by the UK Science and Technology Facilities Council (STFC) via a Centre for Doctoral Training PhD studentship. CA and BL are supported by the European Research Council (ERC) through Starting Grant ERC-StG-716532-PUNCA. BL additionally acknowledges support by the STFC through grants No.~ST/T000244/1 and ST/P000541/1. LH is supported by funding from the ERC under the European Union's Horizon 2020 research and innovation programme grant agreement No.~801781, and by the Swiss National Science Foundation (SNSF) grant 179740. This work used the DiRAC@Durham facility managed by the Institute for Computational Cosmology (ICC), on behalf of the STFC DiRAC HPC Facility (www.dirac.ac.uk). This equipment was funded by BEIS capital funding via STFC capital grants ST/K00042X/1, ST/P002293/1, ST/R002371/1 and ST/S002502/1, Durham University, and STFC operations grant ST/R000832/1. DiRAC is part of the UK National e-Infrastructure.

This work was finalised during the Covid-19 outbreak. The authors would like to thank all essential workers around the world that continue to make huge sacrifices to overcome this pandemic.




\bibliographystyle{JHEP}
\bibliography{bibliography} 

\end{document}